%% file: Cicada-arXiv.tex
\documentclass[10pt,journal,compsoc]{IEEEtran}

\ifCLASSOPTIONcompsoc
\usepackage[nocompress]{cite}
\else
\usepackage{cite}
\fi

\ifCLASSINFOpdf
\else
\fi
\usepackage{cite}
\usepackage{bm}
\usepackage{color}

\usepackage{graphics}
\usepackage[justification=centering]{caption}
\usepackage{subfigure}
\usepackage{float}
\DeclareGraphicsExtensions{.pdf,.jpeg,.png}
\usepackage{cprotect}
\usepackage{algorithm}
\usepackage{algpseudocode}

\usepackage[explicit]{titlesec}
\usepackage{array}
\usepackage{multirow}

\usepackage{url}
\usepackage{footnote}
\usepackage{algorithmicx}
\usepackage{diagbox}

\usepackage{graphicx}
\usepackage{booktabs}
\usepackage{changes}
\usepackage{wrapfig}

\usepackage{chngpage}

\usepackage{amssymb}
\usepackage{pifont}
\usepackage{amssymb}
\usepackage{amsmath,amssymb,amsfonts}
\usepackage{amsmath}
\usepackage{lineno} 
\def\BibTeX{{\rm B\kern-.05em{\sc i\kern-.025em b}\kern-.08em
		T\kern-.1667em\lower.7ex\hbox{E}\kern-.125emX}}

\usepackage{color, xcolor}
\usepackage{soul}

\begin{document}


\title{Cicada: Enabling Pipeline-Efficient Serverless DNN Inference via Decoupled Management}
%
%
%

\author{Zhaorui~Wu,
        Yuhui~Deng,
        Jia~Hu,
        Lin~Cui,
        Zhen~Zhang,
        Lingfang~Zeng,
        Geyong~Min

\IEEEcompsocitemizethanks{\IEEEcompsocthanksitem Z. Wu, Y. Deng, L. Cui, Z. Zhang are with the Department of Computer Science, Jinan University, Guangzhou, Guangdong Province 510632, China. E-mail: diom\_wu@163.com, tyhdeng@jnu.edu.cn, tcuilin@jnu.edu.cn, zzhang@jnu.edu.cn.
\IEEEcompsocthanksitem Z. Wu is also with the Department of Computer Science, University of Exeter, Exeter, EX4 4QF, U.K. E-mail: zw467@exeter.ac.uk.
\IEEEcompsocthanksitem L. Zeng is with the Zhejiang Lab, Hangzhou 311121, China. E-mail: zenglf@zhejianglab.org.
\IEEEcompsocthanksitem J. Hu and G. Min are with the Department of Computer Science, University of Exeter, Exeter, EX4 4QF, U.K. E-mail: g.min@exeter.ac.uk, j.hu@exeter.ac.uk.
}
\thanks{This work has been submitted to the IEEE for possible publication. Copyright may be transferred without notice, after which this version may no longer be accessible.}}

\IEEEtitleabstractindextext{%
	\input{abstract}
	
	}

\maketitle

\IEEEdisplaynontitleabstractindextext

\IEEEpeerreviewmaketitle

\section{Introduction}

Serverless computing represents a significant advancement in cloud resource management, enabling automatic scaling and enhancing cost efficiency for modern applications~\cite{Shafiei2022Serverless, Li2022ServerlessSurvey, Li2023Serverless, Yu2024Freyr}. 
The growing demand for efficient Deep Natural Network (DNN) model deployment across industries, driven by large language models (LLMs) and other foundation models, has accelerated the adoption of serverless architectures for model serving, particularly in scenarios requiring rapid scaling~\cite{Pei2023AsyFunc, Gu2023FaSTGShare, Yang2024VersaDNN}. 
Major cloud providers have integrated this approach into their services, including Amazon SageMaker~\cite{AmazonSageMaker}, Azure Machine Learning~\cite{AzureMachineLearning}, and Google Cloud Vertex AI~\cite{GoogleCloudVertexAI}. 
Recent studies indicate that serverless implementations can lower operational costs compared to traditional cloud services while maintaining performance, particularly for applications with fluctuating workloads~\cite{Wu2022ServerlessDataScience, Yang2022INFless}.

In serverless inference systems, user requests trigger on-demand execution of model-encapsulated functions resident in isolated execution environments, such as containers~\cite{Docker}. 
These operations face a critical performance contrast: container provisioning (cold start) versus reuse of active instances (warm start). 
Empirical studies reveal cold starts incur 5-10x longer latency due to container initialization and model loading processes~\cite{Cai2024Incendio, Pan2022RetentionAware}. 
The cold start challenge in serverless inference systems comprises two phases: (1) function instance startup and (2) model loading. 
Existing optimization approaches have enhanced instance startup efficiency via container management techniques~\cite{Du2020Catalyzer, Zijun2022Pagurus, Roy2022IceBraker, Ashraf2022ORION, Yu2024RainbowCake, Lu2024SMIless}. 
However, optimization of model loading remains limited by preloading~\cite{Sui2024InstaInfer, Pei2023AsyFunc} and layer-sharing methods~\cite{Li2022Tetris, Hong2024Optimus}, as they depend either on precise request prediction mechanisms~\cite{shahrad2020serverless, Golec2024ColdStart} or incur potential memory overhead~\cite{Saxena2022Medes}.



\begin{figure}[t]
    \includegraphics[width=0.48\textwidth]{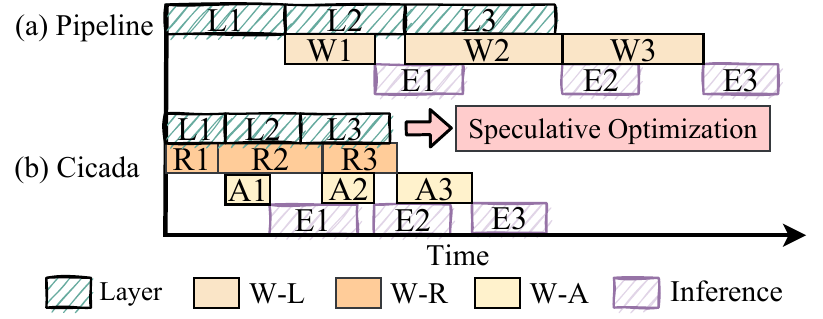}
    \caption{Cicada decouples weight loading (W-L) into weight retrieval (W-R) and weight application (W-A)}
    \label{fig:pipelineModelLoading}
\end{figure}


Building upon the core concept of pipelined execution~\cite{Jin2024DistMind, Ji2022DemandLayering}, the latest research, PISeL~\cite{Rahimi2024PISeL}, introduces pipelined loading mechanism to enhance the efficiency of serverless inference.
As shown in Fig.~\ref{fig:pipelineModelLoading}(a), this approach divides model serving into three layer-wise execution units within a pipeline: (i) layer construction, (ii) weight loading, and (iii) request inference. 
To further improve pipeline efficiency, model loading (stage (i) and (ii)) is executed in parallel with stage (iii).
The feasibility of this pipeline arises from the inherently modular architecture of DNN models, which consist of sequentially dependent computational layers.

Despite the benefits of pipelined execution, there remain inefficiencies in resource utilization. 
Our motivating results (Sec.~\ref{sec:backgroundAndMotivation}) reveal two key inefficiencies in the model inference pipeline. 
First, constructing model layers is intrinsically time-consuming and computationally demanding, as it introduces unnecessary overhead in parameter registration and initialization that prolongs layer instantiation. 
Second, the weight loading process itself is I/O-related, causing execution stalls as the weight loading pipeline unit remains idle while waiting for data transfers to complete. 
This imbalance between computation and data movement exacerbates overall inefficiencies, further limiting pipeline performance. 
Together, these challenges highlight the fundamental bottlenecks in existing inference pipelines and underscore the need for a more efficient execution strategy.

To address these limitations, we propose \textbf{\textit{Cicada}}—a novel pipeline optimization framework that coordinates computational, storage, and scheduling resources (see Fig.~\ref{fig:pipelineModelLoading}(b)).
For layer construction, Cicada implements MiniLoader, which employs a speculative strategy to optimize layer initialization (Speculative Optimizaiton in Fig.~\ref{fig:pipelineModelLoading}(b)), reducing structural preparation time by an average of 40.54\% (detailed in Sec.~\ref{sec:breakdownOfTimeOverheadOfPipelineUnits}).
Aiming to improve the pipeline utilization, Cicada partitions the weight loading unit into two distinct stages: (i) retrieving weight files from disk (W-R in Fig.~\ref{fig:pipelineModelLoading}(b)) and (ii) applying deserialized weight parameters to layer structures (W-A in Fig.~\ref{fig:pipelineModelLoading}(b)). 
Building on this separation, WeightDecoupler is introduced, which presents an asynchronous retrieval strategy to process weight files retrieval before execution, thereby enabling data transfers to overlap with layer construction.
The WeightDecoupler then facilitates out-of-order weight application during the stage (ii), allowing layers to integrate weights as soon as their structures are ready.
Consequently, pipeline stalls are minimized, improving overall efficiency by an average of 42.46\% (see Sec.~\ref{sec:sub:evaluationPipelineEfficiency}).
To manage task execution effectively, Cicada employs a Priority-Aware Scheduler that monitors resource usage across the pipeline and reallocates resources to alleviate contention bottlenecks, thus prioritizing critical inference requests and ensuring stable system performance.
We summarize our contributions as follows.
\begin{itemize}
    \item We identify key performance bottlenecks in pipelined model inference, highlighting inefficiencies in layer construction delays and I/O-induced stalls that create an imbalance between computation and data movement.
    \item  We design Cicada with the following novel optimizations: 
    (1) speculative initialization during layer construction to reduce structural overhead;
    (2) asynchronous weight retrieval combined with out-of-order weight application to eliminate loading stalls. 
    In addition, Cicada employs a priority-aware scheduler to manage task execution effectively, ensuring high-priority inference requests are executed promptly.
    \item We implement a prototype system of Cicada and conduct extensive experiments to validate its effectiveness. The experimental results demonstrate that Cicada achieves up to 64.07\% reduction in end-to-end inference latency compared to the state-of-the-art frameworks, while gains up to 68.49\% improvement in pipeline utilization.
\end{itemize}
\begin{figure}[t]
    \centering
    \includegraphics[width=0.4\textwidth]{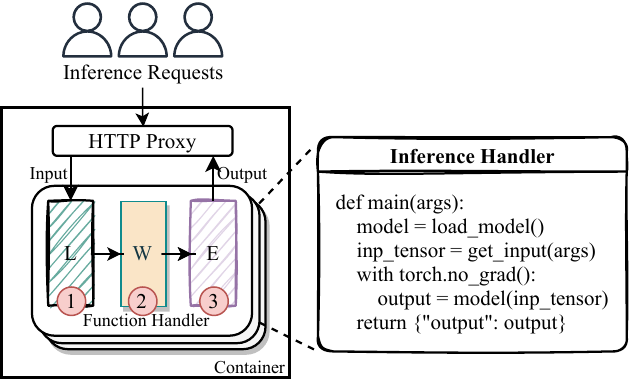}
    \caption{Overview of Serverless Inference}
    \label{fig:serverlessInference}
\end{figure}

\begin{table}[t]
    \centering
    \caption{Model sizes}
    \label{tab:model-sizes}
    \begin{tabular}{cccccc}
        \toprule
        Model     & Size & Model     & Size \\
        \midrule
        ResNet50  & 98 MB        & VGG11     & 506 MB       \\
        ResNet101 & 171 MB       & VGG16     & 527 MB       \\
        ResNet152 & 231 MB       & VGG19     & 548 MB       \\
        \midrule
        LLaMA-3.2-1B  & 4.71 GB       & OPT-1.3B  & 5.09 GB       \\ 
        LLaMA-3.2-3B & 12.26 GB       & OPT-2.7B & 10.12 GB      \\ 
        LLaMA-3-8B & 28.63 GB      & OPT-6.7B & 25.40 GB      \\ 
        \bottomrule
    \end{tabular}
\end{table}

\begin{figure*}[!h]
    \centering
    \begin{minipage}{0.3\textwidth}
        \centering
        \includegraphics[width=\textwidth]{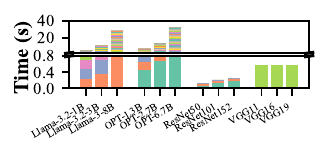}
        \caption{Layer construction overhead}
        \label{fig:layerConstructTimeOverhead}
    \end{minipage}
    \begin{minipage}{0.3\textwidth}
        \centering
        \includegraphics[width=\textwidth]{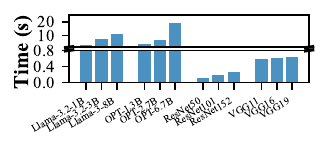}
        \caption{Space allocation overhead}
        \label{fig:layerWeightInitOverhead}
    \end{minipage}
    \begin{minipage}{0.3\textwidth}
        \centering
        \includegraphics[width=\textwidth]{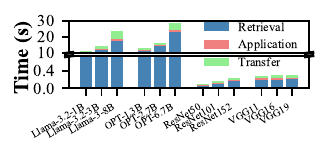}
        \caption{Weight loading overhead}
        \label{fig:layerLoadingAndApplyingOverheadStacked}
    \end{minipage}
\end{figure*}

\section{Background and Motivation}
\label{sec:backgroundAndMotivation}

In this section, we analyze the serverless inference life cycle (Sec.~\ref{sec:serverlessInferenceLifeCycle}), identify the performance bottlenecks in the pipelined model inference (Sec.~\ref{sec:constructionOverhead}), and show the motivation for designing Cicada (Sec.~\ref{sec:opportunityOfDesigningCicada}).

\subsection{Serverless Inference Life Cycle}
\label{sec:serverlessInferenceLifeCycle}
The serverless inference lifecycle comprises three sequential phases: (i) execution environment provisioning if needed (i.e., function cold start), (ii) model loading, and (iii) inference execution.
Following standard serverless paradigms, the provisioning phase establishes an isolated runtime environment with necessary dependencies and a handler function. 

Due to the ephemeral and stateless nature of serverless function execution, model loading and inference execution must be repeated even when requests are routed to warmed instances (i.e., function warm starts). This requirement stems from the container-level process isolation mechanism~\cite{Stojkovic2023MXFaaS, Yu2024RainbowCake}. 
Fig.~\ref{fig:serverlessInference} illustrates the execution process of serverless inference.
A container-embedded HTTP proxy receives user requests and triggers the function handler to initiate inference through the deep learning runtime (e.g., PyTorch, TensorFlow).
The runtime executes the inference task by deserializing persisted model artifacts, following three stages: \textcircled{1} initializing the model structure (L); \textcircled{2} loading weight files and applying parameters to the instantiated model (W); \textcircled{3} performing inference execution (E), where the model processes input data through forward propagation to generate predictions.

Existing research efforts~\cite{Li2022Tetris, Hong2024Optimus} have explored maintaining loaded models in memory to avoid deserialization overhead or sharing models across containers. 
However, these approaches neglect the critical memory pressure imposed by persistent model retention. 
To quantify this overhead, we conduct empirical measurements across four prominent model architectures: ResNet~\cite{He2016ResNet}, VGG~\cite{Karen2015VGG}, LLaMA~\cite{Hugo2023LLaMA} and OPT~\cite{Susan2022OPT}. 
As depicted in Tab.~\ref{tab:model-sizes}, our results demonstrate substantial memory requirements ranging from 98~MB for ResNet50 to 28.63~GB for LLaMA-3-8B. 
This substantial memory footprint renders persistent model retention impractical for production-grade serverless inference systems, particularly under concurrent request handling scenarios.

To improve the efficiency, the latest research PISeL~\cite{Rahimi2024PISeL} introduces pipelined model loading mechanisms to enhance the efficiency of serverless inference. 
Fig.~\ref{fig:pipelineModelLoading}(a) illustrates the model inference process of PISeL.
In this approach, model inference is divided into three layer-wise execution units within a pipeline, including layer construction stage ($L_i$), weight loading stage ($W_i$), and inference execution stage ($E_i$).
With this pipeline design, inference can be executed asynchronously alongside the preceding stages, thereby reducing latency and improving serverless inference throughput. 
For instance, while the inference unit is processing $E_i$, the model loading unit simultaneously executes $L_{i+1}$ and $W_{i+1}$.
However, this pipeline-based model loading approach fails to address the essential problem of pipeline stalls.
The reason behind includes (i) \textbf{the highly time-consuming layer construction process}, which causes the weight loading and inference execution to wait (pipeline stall) for the completion of the layer construction, and (ii) \textbf{the unawareness of computation and data movement imbalance in the weight loading process}, resulting in severe pipeline underutilization.
To further analyze the performance overhead of the pipeline, we conduct a detailed study of the execution overhead.

\subsection{Construction Overhead of Model Layers}
\label{sec:constructionOverhead}

By instrumenting PyTorch, a widely adopted deep learning runtime, we identify two critical phases in layer construction.
While we use PyTorch as our concrete analysis framework due to its popularity and accessibility, the identified bottlenecks and optimization opportunities are inherent to the layer-by-layer construction paradigm shared across all major deep learning frameworks.
The first phase involves the instantiation of layers (e.g., Conv, Linear, Embedding, etc.) to establish the model's computational task, while the second phase entails space allocation through parameter registration and initialization, thereby ensuring designated memory slots for subsequent weight loading.

Although the initial phase defines the computational dependencies of the model, it is the space allocation stage that pre-allocates memory slots, and this stage incurs superfluous overhead in inference scenarios. 
For example, in a 3x3 convolutional layer, parameter registration results in the allocation of thousands of placeholders for weight parameters. 
Moreover, during the parameter initialization phase, these parameters are assigned values using normalization techniques such as the Kaiming normal distribution~\cite{He2015DelvingDeep}. 
However, since pre-trained weights are available during inference, this initialization becomes redundant.

We then illustrate the execution characteristics of four prominent model architectures—ResNet, VGG, LLaMA, and OPT. 
The total layer construction time is determined by the combined duration of both phases, instantiating the computational layers and performing space allocation, as illustrated in Fig.~\ref{fig:layerConstructTimeOverhead} and Fig.~\ref{fig:layerWeightInitOverhead}.
Fig.\ref{fig:layerConstructTimeOverhead} breaks down the construction overhead across different layers (marked by different colors), showing that different layers contribute different overheads to the construction process, ranging from 74.02~ms for ResNet50 to 1150.31~ms for OPT-6.7B.
Besides, Fig.\ref{fig:layerWeightInitOverhead} reveals that the parameter space application phase alone accounts for almost 50\% of the total instantiation (layer construction followed by space allocation) time, despite being unnecessary for inference with pre-trained weights. 
This excessive overhead propagates through the pipeline, delaying weight loading and inference execution. 
Ultimately, these structural inefficiencies undermine the benefits of pipelined execution, as both the weight loading and inference units remain blocked, waiting for layer construction to complete.

In summary, although pipeline-based model loading accelerates execution, inefficiencies in layer construction inherently limit its overall performance.
Therefore, optimizing layer instantiation with a lightweight construction process is crucial to fully unlocking the performance potential of pipelined execution.

\subsection{Opportunity of Designing Cicada}
\label{sec:opportunityOfDesigningCicada}
Inference execution requires pre-trained weights to be loaded into memory, as deep learning models rely on precomputed parameters rather than learning them during inference.
In serverless inference, weights are preferred to be stored on local disks such as SSD or HDD, alongside container images, to avoid additional remote I/O overhead.
Therefore, the weight loading process comprises two distinct phases: (i) weight file fetching and (ii) weight application to the instantiated model.
In the first phase, weight files are accessed from storage, involving operations such as disk retrieval and data deserialization to reconstruct the stored tensors. 
Once the weight data is available in CPU memory, the second phase assigns the deserialized parameters to their corresponding model layers, ensuring proper data alignment.
Subsequently, the fully constructed model is transferred from CPU to GPU memory to prepare for inference execution.

Fig.\ref{fig:layerLoadingAndApplyingOverheadStacked} quantifies the execution overhead of these phases across different model architectures, alongside the transfer time from CPU to GPU memory.
The blue bars represent I/O-bound weight file retrieval, while the pink bars indicate weight application, and the green bars represent the model transfer time.
On average, weight file retrieval, weight application, and weight transfer account for 63.02\%, 9.02\% and 27.96\% of the total loading time, respectively.
Moreover, this disproportionate 7:1 ratio between weight retrieval and application results in severe pipeline underutilization.
Regarding transfer time, it accounts for 6.93\% of the total preparation time, which includes layer construction, space allocation, weight file retrieval, weight application, and model transfer itself.
This relatively small proportion, combined with the distinct staging of retrieval and application phases, suggests that weight loading can be partially overlapped with layer construction and downstream computation.

In addition to the overhead incurred by layer construction (see Fig. \ref{fig:layerConstructTimeOverhead}), the I/O-related nature of weight file retrieval can cause stalls in subsequent pipeline stages, including weight application and inference execution.
The primary cause of this bottleneck is that weight application is contingent upon the complete execution of weight file retrieval, while inference execution demands that the entire layer be fully loaded.
Consequently, it is imperative to devise an efficient strategy that decouples these two operations, thereby mitigating the resultant delays and enhancing overall system performance.
Therefore, we advocate for decoupling weight file retrieval from weight application, allowing retrieval to overlap with layer construction. 
This separation reduces pipeline stalls by breaking the sequential dependency between retrieval and application, enabling independent scheduling of weight application tasks.
Furthermore, this decoupling also creates an opportunity for out-of-order execution of weight application, without waiting for the current weight file retrieval to complete, alleviates stalls caused by sequential dependencies and thereby enhances overall pipeline efficiency.

However, it is important to note that this decoupling introduces a potential instability issue related to the scheduling of weight retrieval from a performance stability perspective (detailed in Sec.~\ref{sec:priorityAwareScheduler}). 
To ensure performance stability, we propose a priority-aware scheduler that assigns precedence to inference execution units based on workload demand.

\begin{figure}[t]
    \raggedleft
    \includegraphics[width=0.48\textwidth]{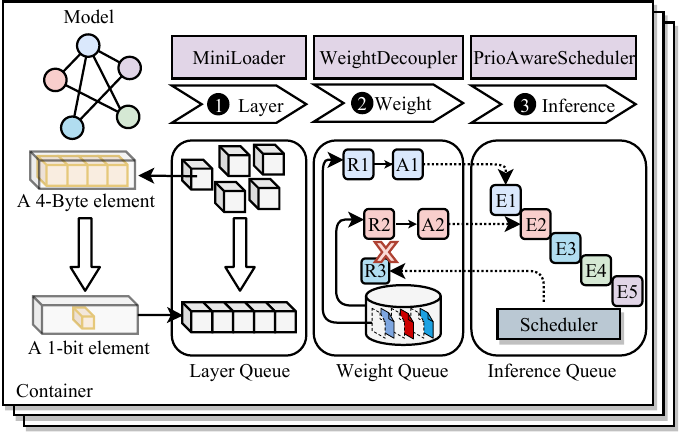}
    \caption{Cicada Overview}
    \label{fig:CicadaOverview}
\end{figure}

\section{Design}
\label{sec:design}

\subsection{Design Overview}
In this section, we introduce Cicada, a novel pipeline optimization framework that effectively coordinates computational, storage, and scheduling resources. 
As with other serverless inference systems, users submit inference requests (e.g., the desired model, weight, and input data) to the serverless platform, which subsequently invokes the Cicada framework to process the request.
Embedded within the serverless runtime (e.g., a container), Cicada comprises three primary components (shown in Fig.~\ref{fig:CicadaOverview}):

\begin{enumerate}
    \item \textbf{\textit{MiniLoader}}: This module applies a speculative strategy to dynamically adjust parameters registration and initialization to minimize unnecessary overhead in layer construction phase. 
    \item \textbf{\textit{WeightDecoupler}}: To accelerate weight processing, Cicada decouples weight file retrieval from weight application, allowing its overlap with layer construction.
    This separation permits out-of-order weight application ($A_1, A_2$) by executing weight file retrieval ($R_1, R_2$) asynchronously alongside layer construction.
    \item \textbf{\textit{Priority-Aware Scheduler}}: In the inference execution phase, Cicada employs a Priority-Aware Scheduling mechanism, dynamically prioritizing inference execution units ($E_1, E_2$) based on workload demand.
\end{enumerate}

Upon receiving an inference request, Cicada initiates layer construction, weight loading, and inference execution in a layer-by-layer manner.
During this process, the \textbf{\textit{MiniLoader}} identifies space allocation operations and speculatively substitutes the corresponding parameters to reduce construction overhead and temporarily alleviate memory stress. 
Furthermore, the \textbf{\textit{WeightDecoupler}} employs a decoupling strategy for weight file retrieval by overlapping it with layer construction. 
This decoupling allows weight file retrieval to proceed independently of layer construction, thereby enabling weight application to be executed out of order.
Finally, by utilizing the \textbf{\textit{Priority-Aware Scheduler}}, Cicada guarantees that inference execution is assigned the highest priority. This ensures that critical tasks are executed promptly, thereby optimizing overall pipeline efficiency.

\subsection{MiniLoader}
\label{sec:miniLoader}
As detailed in Sec.~\ref{sec:constructionOverhead}, the parameter registration and initialization phase during layer construction incurs superfluous overhead in inference scenarios. 
Conventionally, the registration process allocates full-precision (typically float32, or 4 bytes) placeholders according to each layer's dimensions (e.g., $in\_channels$, $out\_channels$, $kernel\_size$, etc.)
This approach, while ensuring numerical accuracy and preserving the integrity of the layer's computation graph, results in memory overhead during model construction. 
For instance, a layer may instantiate a \texttt{weight} parameter by using \texttt{nn.Parameter(torch.empty(x, y, z))}, which creates a tensor with $x \times y \times z$ 4-byte placeholders.
However, the actual numerical values in this tensor are temporary, as they will be overwritten by certain initialization methods or pre-trained weights during inference.

\subsubsection{\textbf{Parameters Registration}}

Given that these initial parameter values are ultimately replaced by pre-trained weights, it is feasible to utilize a low-precision format without compromising the structural requirements. 
To that end, Cicada introduces MiniLoader, an speculative adjustment mechanism that reduces both initialization latency and memory footprint. 
Specifically, MiniLoader takes control of the parameter initialization process by employing low-precision techniques that compress the conventional 4-byte (32-bit) placeholders into 1-bit representations (see the lower left panel of Fig.~\ref{fig:CicadaOverview}). 
This substitution can significantly reduce 32x memory consumption and 31.88\% initialization overhead (see Sec.~\ref{sec:experimentResults}) without compromising numerical correctness.
In practice, monitoring the parameter construction process involves intercepting the allocation and registration of parameter objects (for instance, via PyTorch's internal calls to \texttt{register\_parameter} or \texttt{Parameter.new}), and then applying custom logic to convert placeholder-only parameters to a compressed, low-precision format.
When weight loading mechanism (e.g., \texttt{load\_state\_dict}) is invoked, the pre-trained weights are then copied into these objects, restoring the necessary precision for accurate inference.

\subsubsection{\textbf{Parameters Initialization}}
In inference scenarios, the availability of pre-trained weights obviates the need for subsequent parameter initialization. 
While the parameter initialization phase, often executed via sophisticated schemes such as Kaiming initialization~\cite{He2015DelvingDeep}, remains essential for establishing the layer's computation graph during training, its outcomes are ultimately superseded by the loaded weights during inference.
Recognizing that parameter initialization is redundant during inference, MiniLoader bypasses these superfluous steps entirely by skipping the default initialization process, thereby halving the overall initialization time (comparing Fig.\ref{fig:layerConstructTimeOverhead} and Fig.\ref{fig:layerWeightInitOverhead}).

\begin{figure}[t]
    \includegraphics[width=0.5\textwidth]{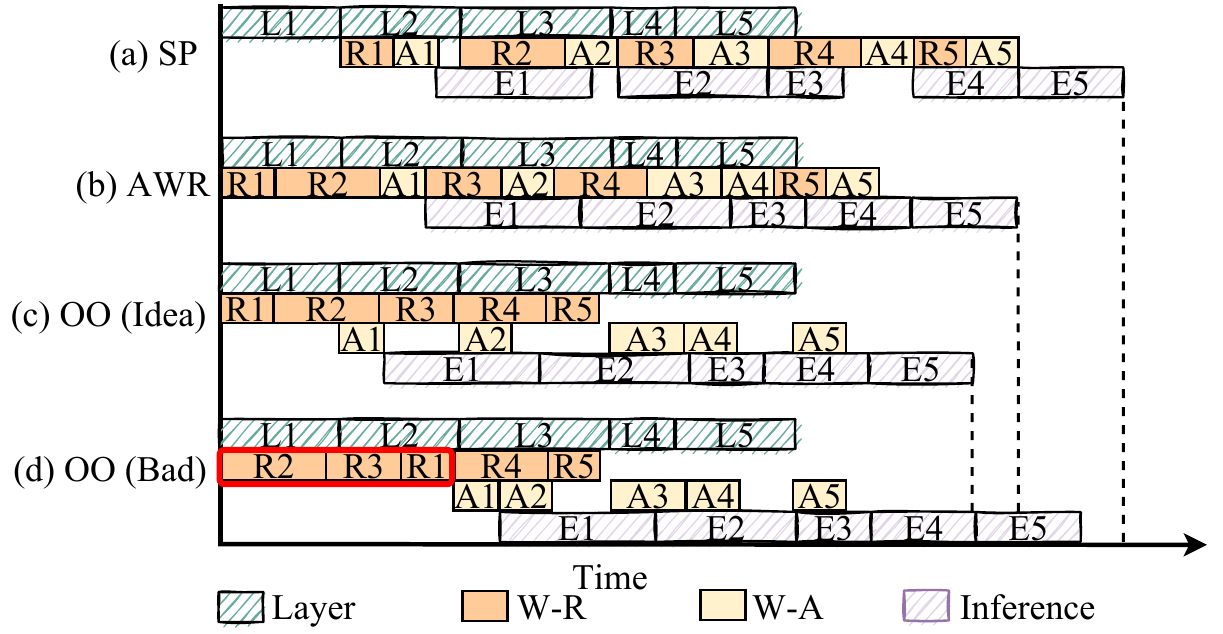}
    \caption{Pipeline Scheduling Strategies in a Five-layer Model}
    \label{fig:pipelineDesign}
\end{figure}

\subsection{Weight Decoupler}
\label{sec:weightDecoupler}
We now turn our attention to optimizing the weight loading process with the Weight Decoupler of Cicada. 
In this process, the deep learning runtime (e.g., PyTorch, TensorFlow) is responsible for applying pre-trained weights to an instantiated model.
Typically, pre-trained weights are stored as weight files (\texttt{.pth} for PyTorch and \texttt{.pb} for TensorFlow), which must be retrieved from storage and loaded into memory before being applied to the model. 
This weight loading process can be divided into two sequencial stages: (1) weight file retrieval which is I/O-bound and (2) weight application which is computation-bound.
Fig.~\ref{fig:pipelineDesign}(a) illustrates a sequential pipeline-based (SP) inference process of a 5-layer model, where the weight loading process ($R_i$ and $A_i$) is sequentially dependent on the layer construction process ($L_i$), leading to inefficiencies caused by blocked execution.

Fortunately, layer construction and weight file retrieval operate independently.
Exploiting this natural separation, the Weight Decoupler enables Asynchronous Weight Retreival (AWR) by overlapping weight file fetching with layer construction, with the aim of improving pipeline efficiency.
As shown in Fig.~\ref{fig:pipelineDesign}(b), weight file operation ($R_1$) is processed concurrently during the construction of layer 1 ($L_1$).
Once both $R_1$ and $L_1$ complete, the weight application ($A_1$) is performed, allowing inference of layer 1 ($E_1$) to proceed.
The feasibility of this decoupling is supported by the operation overhead analysis in Fig.~\ref{fig:layerConstructTimeOverhead} - Fig.~\ref{fig:layerLoadingAndApplyingOverheadStacked}, which shows that weight loading (including weight file retrieval and weight application) is lighter compared to layer construction.

To implement the AWR mechanism, the Weight Decoupler launches a dedicated I/O process in the backend to continuously monitor and handle weight file retrieval requests asynchronously. 
However, memory management is critical during weight retrieval since buffering weight files for large models can exhaust available resources (see Tab.~\ref{tab:model-sizes}). 
To mitigate this, we introduce a per-layer memory management system in which memory pages for each layer are released at the end of every inference. 
This strategy is supported by the observation that layer construction takes longer than weight retrieval, thereby enabling the pipeline to effectively overlap the weight loading overhead. 
By releasing memory after each inference, the system promptly clears unused data and maintains optimal memory utilization even when handling inferences.

In a nutshell, Cicada's decoupling strategy overlaps weight file retrieval with layer construction within an active pipeline, improving concurrency between I/O and compute stages while managing memory efficiently by releasing per-layer buffers after each inference.

\subsection{Out-of-Order Weight Application}
\label{sec:outOfOrderWeightApplication}
AWR ensures that weight files are fetched concurrently from storage while layers are being instantiated, reducing idle waiting time and enhancing overall pipeline throughput. 
In practice, while weight file retrieval handles I/O tasks, the weight loading unit remains idle because the kernel autonomously transfers data from storage to memory once an I/O request is issued. 
This creates a distinct asynchronous gap during which the weight loading unit is not engaged in I/O processing, and therefore, can be repurposed to perform weight application for the preceding layer. 
As a result, once a layer's construction is complete, the weight loading unit can immediately apply its corresponding weights, regardless of ongoing weight retrieval for subsequent layers, thus permitting out-of-order (OO) execution. 

However, similar to dependency hazards in CPUs~\cite{Sharafeddine2012Disjoint}, enabling out-of-order execution introduces the challenge of preventing data hazards while maintaining the correct operational sequence: layer construction $\rightarrow$ weight retrieval $\rightarrow$ weight application. 
To support out-of-order execution without violating these dependencies, our design ensures that the weight application for each layer is triggered only after three conditions are met: (i) the weight data for that layer is fully retrieved, (ii) the construction of the current layer is complete, and (iii) the weight application for the preceding layer has finished.
Specifically, upon completion of $R_i$, the weight loading unit dispatches a signal to a dedicated task queue, flagging that the weight data is ready. 
Then, Cicada verifies that $L_i$ has been fully constructed.
Moreover, before executing $A_i$, Cicada verifies that $A_{i-1}$ has finished.
During this dependency verification process, Cicada employs fine-grained locking and atomic counters to prevent race conditions during state updates, ensuring robust and consistent execution throughout the pipeline.

Overall, by overlapping asynchronous weight retrieval with layer construction and enforcing a readiness-driven out-of-order execution policy with robust dependency verification, fine-grained locking, and atomic counters, Cicada maximizes pipeline parallelism and minimizes idle stages throughout the model loading path, ensuring a consistent and race-free initialization process.

\subsection{Priority-Aware Scheduler}
\label{sec:priorityAwareScheduler}
The out-of-order mechanism may sometimes suffer from instability, as weight file retrieval operations might not return in the same order in which they were requested because of unpredictable I/O completion.
Due to the inherent complexity of I/O schedulers~\cite{Aupy2019IOSurvey, Sa2018SharedDisk, Chen2016InternalParallelism} and storage media (e.g., SSDs, HDDs), asynchronous I/O requests cannot guarantee a specific return order.
For example, as the red rectaggle shown in Fig.~\ref{fig:pipelineDesign}(d), even if the request for $R_1$ is issued before those $R_2$ and $R_3$ (with the order illustrated in Fig.~\ref{fig:pipelineDesign}(c)), the response for $R_1$ may still be delayed relative to the others. 
This uncertainty in asynchronous I/O may compromise pipeline stability, particularly by affecting the timing of weight application.

To address this issue, we employ a priority-aware scheduler to prioritize inference execution units based on workload demand.
This mechanism ensures that high-priority inference requests are executed promptly while optimizing pipeline efficiency.
This strategy schedules tasks by monitoring the execution status of individual pipeline units. 
Specifically, it adjusts the weight file retrieval processes based on the current progress of the layers. 
The scheduler tracks each layer's ($L$) and weight retrieval's ($R$) execution times, selectively blocking lower-priority I/O to ensure timely completion of the most required weight. 
For example, if layer $L_1$ begins at time $t_0$ with duration $D_{L_1}$, and weight retrieval $R_1$, $R_2$, and $R_3$ start sequentially at $(t_0 + a)$ (each with its own I/O duration, $D_{R_1}$, $D_{R_2}$, $D_{R_3}$), the scheduler checks whether $R_1$ will finish by $(t_0 + a) + D_{R_1}$. If $R_1$ is still running past this expected completion time, the scheduler suspends $R_2$ and $R_3$ to prioritize $R_1$.
The pseudocode of this strategy is shown in Algorithm~\ref{alg:weight_priority_li}.
In the worst-case scenario, where all $N$ concurrent weight operations require suspension, the algorithm exhibits a linear time complexity of $O(N)$, which is effectively constant as $N$ is typically relatively small (e.g., ResNet, VGG, LLaMA, and OPT consist of up to 10, 5, 34, and 35 layers, respectively).

In addition, although this is unlikely due to the dual safeguards provided by dependency verification (Sec.~\ref{sec:outOfOrderWeightApplication}) and priority-based scheduling, to account for potential inaccuracies in the recorded duration \(D_{R_i}\) caused by resource contention, the scheduler proactively prioritizes the weight retrieval operation \(R_i\) whenever the pipeline unit is stalled waiting to execute stage \(A_i\).
This helps speed up its execution and reduce any resulting delay.


\begin{algorithm}[t]
    \caption{Adjust Retrieval Operation Priority for \(L_i\)}
    \label{alg:weight_priority_li}
    \begin{algorithmic}[1]
    \Require 
      \Statex \textbf{$t_0$}: Start time of layer \(L_i\); \textbf{$D_{R_i}$}: I/O duration for \(R_i\)
      \Statex \textbf{$\mathcal{R}$}: Retrieval operations; \textbf{$t$}: Current time
    \Ensure Priority status for \(R_i\)
    
    \State \(expected\_completion \gets (t_0 + a) + D_{R_i}\)
    \If {\(t \geq expected\_completion\) \textbf{and} \(R_i\) is not completed}
        \For {each retrieval operation \(R \in \mathcal{R}\) with \(R \neq R_i\)}
             \State \textbf{block} \(R\) \Comment{Suspend lower-priority  I/O}
        \EndFor
        \State \textbf{set} priority of \(R_i\) to \texttt{HIGH}
    \EndIf
    \State \Return priority of \(R_i\)
    \end{algorithmic}
\end{algorithm}

\section{Evaluation Environment and Methodology}

\subsection{Experiment Setup and Baselines}
This study focuses on the performance and efficiency of Cicada running on a single-node server, rather than a efficiency of clustered server.
To this end, we develope and evaluate Cicada on our own Ubuntu 22.04 server environment, which is equipped with a NVIDIA GeForce RTX 3090 GPU, 2 Intel(R) Xeon(R) Gold 6248R CPUs, 512GB DDR4 memory and Samsung SSD 970 EVO Plus 2TB.
Since this study mainly focuses on optimizing the inference performance in serverless environment, all invocations are dispathced after containers are launched.
Therefore, the cold start latency of containers is not considered in this study.

We compare Cicada with three state-of-the-art frameworks: PISeL~\cite{Rahimi2024PISeL}, a pipeline-based inference system, Tetris~\cite{Li2022Tetris}, which leverages an in-memory tensor caching mechanism to accelerate inference; and Optimus~\cite{Hong2024Optimus}, a transformation-driven inference system that reuses existing models in memory to serve structurally similar target models.

Additionally, we evaluate different component activation strategies within Cicada, specifically, Mini, Preload, and full Cicada (Mini + Preload) configurations.
The Mini strategy incorporates the MiniLoader component (sec.~\ref{sec:miniLoader}) into the PISeL framework, while the Preload strategy integrates the WeightDecoupler component (sec.~\ref{sec:weightDecoupler}) on top of the PISeL framework.
Notably, the Priority-Aware Scheduler (sec.~\ref{sec:priorityAwareScheduler}) is employed in both the Preload and Cicada strategies to mitigate potential performance instability arising from the out-of-order weight application in the WeightDecoupler component, whereas the Mini strategy does not require this component.

\begin{figure}[t]
    \centering
    \includegraphics[width=0.4\textwidth]{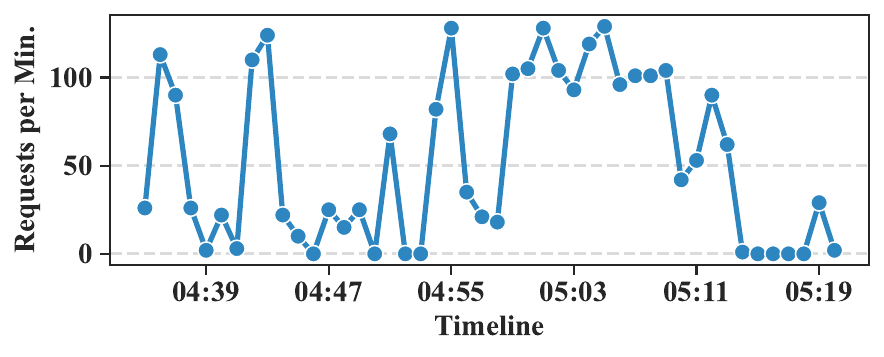}
    \caption{Invocation Pattern}
    \label{fig:invocationPattern}
\end{figure}

\begin{figure*}[!ht]
    \centering
    \minipage{0.45\textwidth}
        \includegraphics[width=1\textwidth]{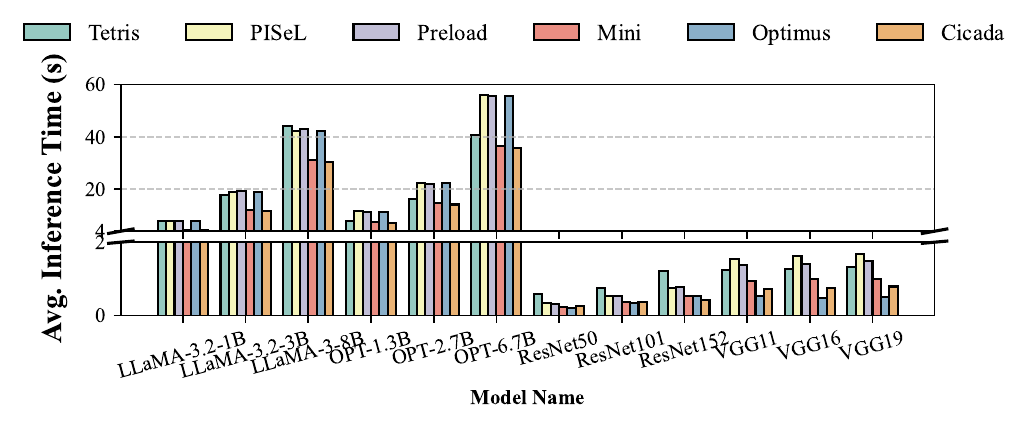}
        \caption{End-to-end inference latency}
        \label{fig:inferenceOverhead}
    \endminipage
    \hspace{10pt}
    \minipage{0.5\textwidth}
        \includegraphics[width=1\textwidth]{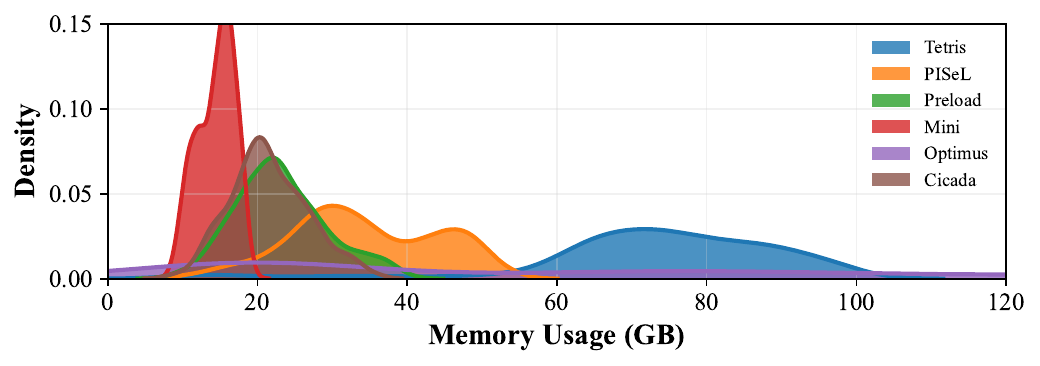}
        \caption{Memory usage distribution}
        \label{fig:resourceUtilizationDistribution}
    \endminipage
\end{figure*}
\subsection{Workloads}
We evaluate Cicada's performance across a diverse set of workloads, including inference requests from the ResNet family (ResNet50, ResNet101, ResNet152), the VGG family (VGG11, VGG13, VGG16, VGG19), the LLaMA family (LLaMA-3.2-1B, LLaMA-3.2-3B, LLaMA-3-8B) and the OPT family (OPT-1.3B, OPT-2.7B, OPT-6.7B).
The primary goal of this study is to optimize the processes of model construction and weight loading (i.e., weight file retrieval and weight application), excluding the inference execution phase. 
To accommodate the heterogeneous input requirements of different model families, we generate synthetic test tensors with model-specific shapes and value distributions.
For language models such as LLaMA and OPT, the input is a 2-dimensional tensor representing token sequences.
Specifically, LLaMA models use integer tensors of shape \texttt{1x128} sampled from a vocabulary of size 32K, reflecting typical autoregressive input patterns.
OPT models are given input sequences of shape \texttt{1x10}, with token values sampled from a vocabulary of size 1K.
For vision models such as ResNet and VGG, the input is a 4-dimensional floating-point tensor of shape \texttt{1x3x224x224}, simulating a single RGB image at the standard ImageNet resolution~\cite{He2016ResNet, Deng2009ImageNet}.

Since Cicada is designed to optimize the process of serverless inference requests, we drive these workloads using the Azure function trace~\cite{zhang2021Faster}. The Azure trace comprises 1,980,951 function execution times and invocation timestamps collected over a 14-day period. 
For our experiments, we utilize data from day 13, randomly assigning functions to the evaluated models over a one-hour duration, resulting in a total of 2426 invocations.
Fig.~\ref{fig:invocationPattern} shows the per-minute invocation pattern, exemplifying the burstiness of serverless functions.
\subsection{Evaluation Metrics}
\label{sec:evaluationMetrics}
To evaluate the effectiveness of Cicada, we aim to address the following key questions:

\textit{\textbf{Q1: How do Cicada perform in terms of inference latency?}} 

In this study, we define \textbf{inference latency} as the total duration from when the serverless function receives an inference request until the response is fully completed.
This metric evaluates the overall performance of Cicada in terms of user experience.
Note that this metric does not include the time for container initialization.

\textit{\textbf{Q2: What memory advantages does the Cicada offer?}}

As discussed in Sec.~\ref{sec:miniLoader}, the MiniLoader component reduces memory overhead during the interval between layer construction and weight loading by optimizing the parameter registration and initialization phases.
Since weight application assigns pre-trained weights to model parameters, the memory reduction achieved by MiniLoader persists until the weight application is complete.
In addition, the WeightDecoupler component reduces memory overhead by explicitly releasing memory pages associated with each layer at the end of every inference.
To quantify \textbf{memory usage} under different strategies, we record system memory consumption using the \texttt{free} command at a sampling frequency of 1 second during execution.


\textit{\textbf{Q3: What is the time overhead within each pipeline units?}} 

An inference pipeline consists of a sequence of pipeline units, including layer construction, weight loading, and inference execution. In this context, we focus on the time overhead incurred by each pipeline unit, which comprises both \textbf{working time} and \textbf{waiting time}. 
The working time of a unit is determined by sampling its start and end timestamps, while the waiting time is calculated as the difference between the start timestamp of the current unit and the end timestamp of the previous unit.

\textit{\textbf{Q4: How efficient is Cicada's pipeline?}}

The execution processes of the various pipeline units overlap. 
For instance, the inference phase of layer $i$ can run concurrently with the construction phase of layer $i+1$.
Thus, \textbf{pipeline utilization} and \textbf{visualization of pipeline timeline} (Gantt chart) are introduced to evaluate the efficiency and parallelism of resource scheduling within the inference pipeline. 
Pipeline utilization is calculated as the ratio of the \textbf{total active time} to the \textbf{total pipeline time}. 
The active time is determined by aggregating the time intervals associated with individual pipeline stages after merging any overlapping periods to ensure that concurrent operations are not counted multiple times.

\section{Experiment Results}
\label{sec:experimentResults}

\subsection{Evaluating Overall Inference Latency}
\label{sec:evaluatingOverallInferenceLatency}
This section details the measurement of overall inference latency, defined as the total time from inference request arrival at the serverless function to response completion. 

As shown in Fig.~\ref{fig:inferenceOverhead}, the Preload, Mini, and Cicada strategies yield various degrees of performance enhancement relative to PISeL, Tetris and Optimus. 
Overall, the Preload, Mini, and Cicada strategies reduce latency by 12.79\%, 34.96\%, and 39.89\% compared to PISeL, while, relative to Tetris, Cicada and Mini achieve improvements of 36.56\% and 30.74\%, respectively, with Preload matching Tetris's performance.
Compared to Optimus, Cicada achieves an average performance improvement of 36.28\% across the LLaMA and OPT families, while Mini delivers a 34.42\% gain over Optimus on the same set of models.
The fundamental reason for Optimus's lower performance lies in its handling of large language models.
Due to their complex and deeply layered architectures, transforming such models at the layer granularity becomes significantly more difficult and costly.
Finding transformation more costly than cold-starting, Optimus reverts to a traditional cold-start for each inference, leading to notable latency and reduced efficiency.
Regarding vision models (ResNet and VGG), Cicada performs comparably to Optimus on ResNet, while Optimus demonstrates a 33.95\% improvement on VGG.
This performance gain stems from Optimus's ability to transform a target model from a source model already resident in memory—for example, transforming VGG19 by leveraging a preloaded VGG11.
This approach effectively trades memory usage for reduced loading overhead.
We provide a detailed discussion of the memory overhead in Sec.~\ref{sec:evaluatingMemoryOverheadOfMiniLoader}.

Above results indicate that the MiniLoader component plays a significant role, suggesting that the models experience substantial unnecessary parameter registration and initialization (see Sec.~\ref{sec:constructionOverhead}).
Specifically, when comparing Cicada to Preload, the VGG family exhibits the most substantial relative performance gains from the MiniLoader component, with improvements ranging from 46.34\% (VGG19) to 47.34\% (VGG11).
In terms of absolute latency reduction, OPT-6.7B benefits the most, with a reduction of 19.83 s, followed by LLaMA-3-8B at 12.62 s.
On average, MiniLoader contributes a 38.32\% speedup across all evaluated models.
Conversely, when comparing Cicada to Mini, the additional improvements brought by the Preload (WeightDecoupler) component are more modest, averaging 7.87\% across models.
The largest relative gains are observed in the VGG family, with VGG16 achieving the highest improvement at 23.87\%.
In terms of absolute latency reduction, the highest gain again comes from OPT-6.7B, with a reduction of 838.28 ms.

Overall, among the different Cicada implementations, the optimization brought by MiniLoader is more pronounced, while the WeightDecoupler module further refines the weight loading phase.
This outcome is largely attributable to the fact that the optimizable overheads in both layer construction and weight file retrieval offer significant opportunities for performance improvement (shown in Fig.~\ref{fig:layerWeightInitOverhead} and Fig.~\ref{fig:layerLoadingAndApplyingOverheadStacked}).

\subsection{Evaluating Memory Overhead}
\label{sec:evaluatingMemoryOverheadOfMiniLoader}
Here, we assess the memory overhead associated with different strategies.
In Cicada, the MiniLoader reduces memory overhead during the interval between layer construction and weight loading.
In Preload, where the MiniLoader is not used, memory overhead is mitigated by explicitly releasing memory pages associated with each layer at the end of every inference.
In Tetris, tensors from different layers are cached in memory and shared across inference requests, enabling reuse and reducing redundant computations.
Optimus, by contrast, stores entire source models in memory and performs fine-grained, layer-level transformations to instantiate structurally related target models at runtime.

Fig.~\ref{fig:resourceUtilizationDistribution} presents the memory usage distribution across these strategies.
From the results, we observe that the Mini strategy exhibits a substantial reduction compared to other strategies.
Mini's memory consumption consistently maintains below 20 GB, while other strategies' memory consumption achieves up to 40 GB (Preload and Cicada), 60 GB (PISeL), 100 GB (Tetris) and even 120 GB (Optimus).
More specifically, Cicada achieves a 45.77\% reduction compared to PISeL, saving 11.6 GB of memory on average.
Against Tetris, Cicada reduces memory usage by 78.82\%, corresponding to a saving of 51.1 GB.
When compared to Optimus, Cicada achieves the highest reduction of 81.82\%, saving 61.9 GB of memory.

This reduction is attributed to MiniLoader replacing PISeL's default float32 (32-bit) precision with 1-bit precision data during the layer construction phase.
This approach is feasible because these values function solely as placeholders between layer construction and before weight application.  
Crucially, MiniLoader restores the parameters to their default precision before weight application to ensure accurate inference.
In addition, the WeightDecoupler component reduces memory usage by freeing per-layer memory immediately after inference.

\begin{figure}[t]
    \centering
    \includegraphics[width=0.35\textwidth]{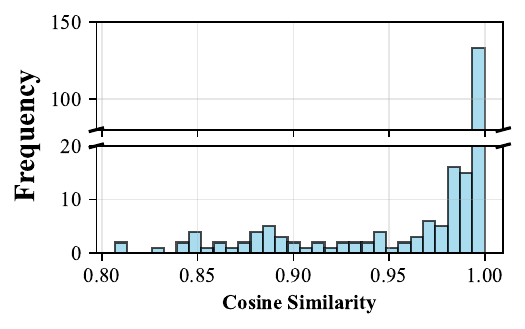}
    \caption{Cosine similarity of Mini and PISeL}
    \label{fig:similarity_distributions}
\end{figure}

To quantify the feasibility of MiniLoader during the subsittution phase, we compare the consine similarity between intermediate layer outputs produced by Mini and PISeL under identical models and inputs.
This evaluation allow us to validate the correctness of MiniLoader's lightweight parameter representation.
Fig.~\ref{fig:similarity_distributions} presents the cosine similarity distributions between Mini and PISeL across all evaluated models, comprising a total of 224 layers.
Among these, 144 layers exhibit a similarity greater than 0.99, and 54 layers fall within the range of [0.95, 0.99).
The average similarity across all layers is 0.974, demonstrating that MiniLoader closely approximates PISeL's activations in the vast majority of cases. These results validate the correctness and functional equivalence of MiniLoader during the layer construction phase.

In summary, the combination of MiniLoader and WeightDecoupler significantly reduces memory overhead during inference, both by minimizing intermediate allocations and reclaiming memory promptly after execution.

\begin{figure*}[t]
    \includegraphics[width=1\textwidth]{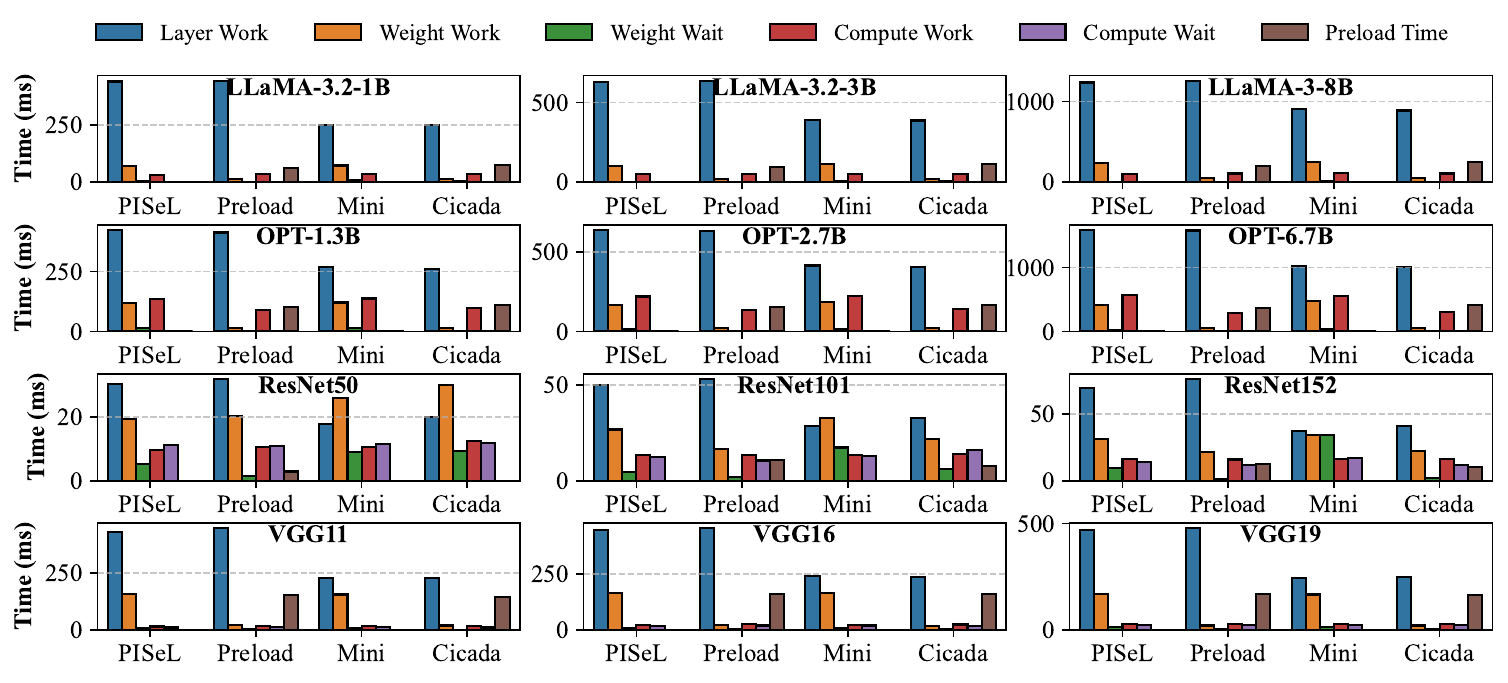}
    \caption{Metrics breakdown for different models across strategies}
    \label{fig:metricsBreakdownGrid}
\end{figure*}

\subsection{Breakdown of Time Overhead of Pipeline Units}
\label{sec:breakdownOfTimeOverheadOfPipelineUnits}
In this section, we  provide a detailed breakdown of the time overhead incurred by individual pipeline units, including Layer, Weight, and Compute. 
This overhead is divided into two components: \textbf{working time} and \textbf{waiting time}. 
Working time denotes the duration during which a unit actively performs its designated tasks, whereas waiting time represents the period during which the unit remains idle while waiting for the completion of computations from the preceding unit.

Fig.~\ref{fig:metricsBreakdownGrid} provides a comprehensive overview of the time overhead for each pipeline unit across different models and strategies. 
Specifically, it presents the average working time for each pipeline unit.
It is noteworthy that the waiting time for the Layer unit (Layer Wait), which represents the interval from when the function receives the request to the start of layer construction (associated with pipeline resource initialization), is negligible. 
As a result, this phase is not shown in the figure.

From Fig.~\ref{fig:metricsBreakdownGrid}, we can observe that, in PISeL, the working time of the Layer unit is significantly higher than that of the Weight unit, which is consistent with the findings in Sec.~\ref{sec:constructionOverhead}.
PISeL and the Preload strategy achieve similar layer construction overhead (Layer Work); however, the Preload strategy yields an average of 69.45\% improvement in Weight Work compared to PISeL across all models, specifically, ranging from 32.33\% (ResNet152) to 88.77\% (OPT-1.3B).
This improvement is attributed to the WeightDecoupler, which splits the weight loading phase into weight file retrieval and weight applying, thereby shifting the file retrieval overhead into the Preload Time (represented by the brown bars in the figure). 
Note that the file retrieval in WeightDecoupler overlaps with the layer construction process, so this preload time does not result in any additional time overhead.

The Mini strategy, compared to PISeL, demonstrates a clear improvement in the Layer Work phase, achieving an average optimization of 40.54\%, with individual improvements ranging from 26.89\% (LLaMA-3-8B) to 47.68\% (VGG19) across different models.
Above improvements are attributed to the MiniLoader, which reduces the layer construction overhead by optimizing the parameter registration and initialization phases.

\begin{figure*}[t]
    \minipage{0.48\textwidth}
        \includegraphics[width=1\textwidth]{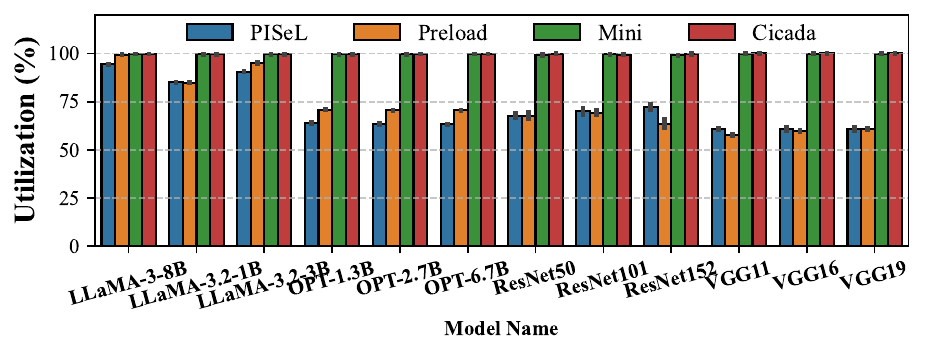}
        \caption{Pipeline utilization}
        \label{fig:pipelineUtilization}
    \endminipage
    \hspace{10pt}
    \minipage{0.48\textwidth}
        \includegraphics[width=1\textwidth]{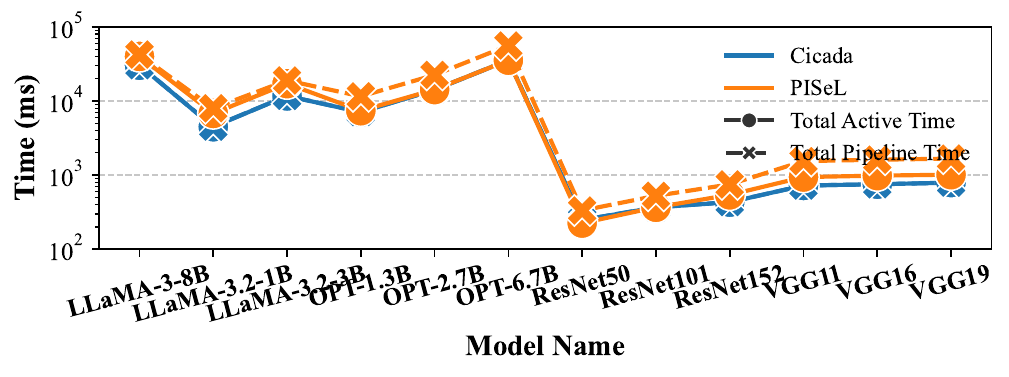}
        \caption{Pipeline times comparison}
        \label{fig:pipelineTimesComparison}
    \endminipage
\end{figure*}

Delving deeper into the pipeline, we observe that Cicada and Mini exhibit prolonged waiting phases in Resnet family.
Notably, the MiniLoader component increases both the Weight Wait and Compute Wait durations, whereas the WeightDecoupler primarily extends the Compute Wait phase.
Under the Mini strategy, the Weight Wait phase exhibits an average degradation of 49.35\%, peaking at 74.98\% (ResNet101), whereas the Compute Wait phase shows an average degradation of 7.25\%, reaching up to 16.81\% (ResNet152).
The underlying reason is that MiniLoader accelerates the layer construction process, causing the Weight Loading component to receive more requests per unit time, which in turn increases the waiting time for weight loading tasks.
Furthermore, as the downstream stages cannot process these additional requests quickly enough, the Compute Wait phase is also prolonged, albeit to a lesser extent than the Weight Wait phase. 

Please note that pipeline units continue executing other requests even while a particular one experiences Weight Wait or Compute Wait. 
Therefore, the additional waiting overhead is deemed acceptable in light of the overall performance improvements.
Notably, the observed waiting is not caused by any implementation inefficiency introduced by MiniLoader. Rather, it results from a relative increase in the proportion of waiting time due to the accelerated layer construction. In absolute terms, Cicada and its variants achieve lower end-to-end latency, underscoring the effectiveness of the proposed optimizations.

\begin{figure*}[t]
    \centering
    \subfigure[ResNet152]{
        \includegraphics[width=0.3\textwidth]{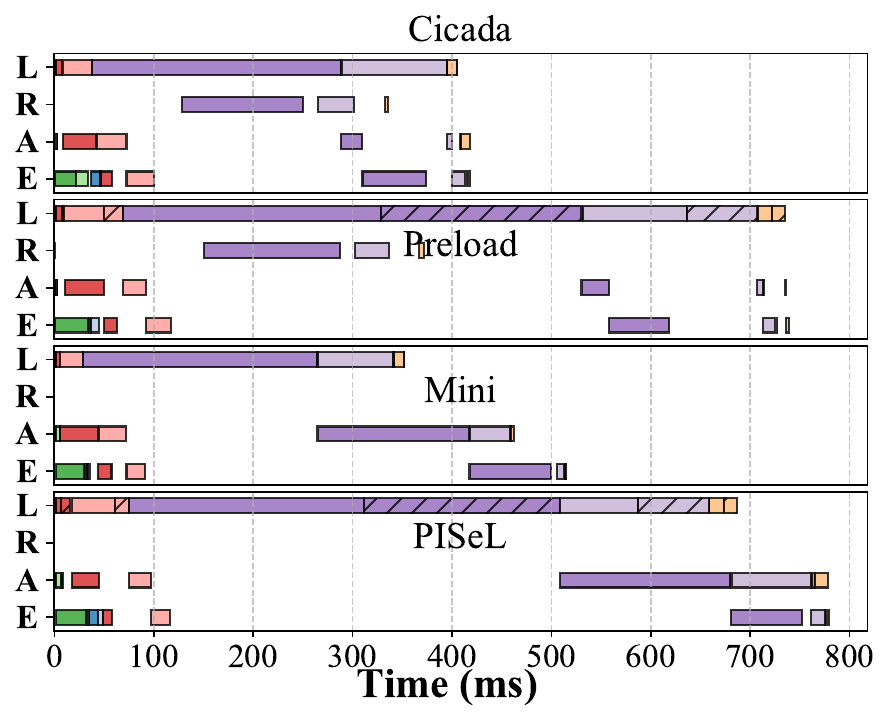}
        \label{fig:pipelineTimelineResnet152}
    }
    \subfigure[LLaMA-3-8B]{
        \includegraphics[width=0.3\textwidth]{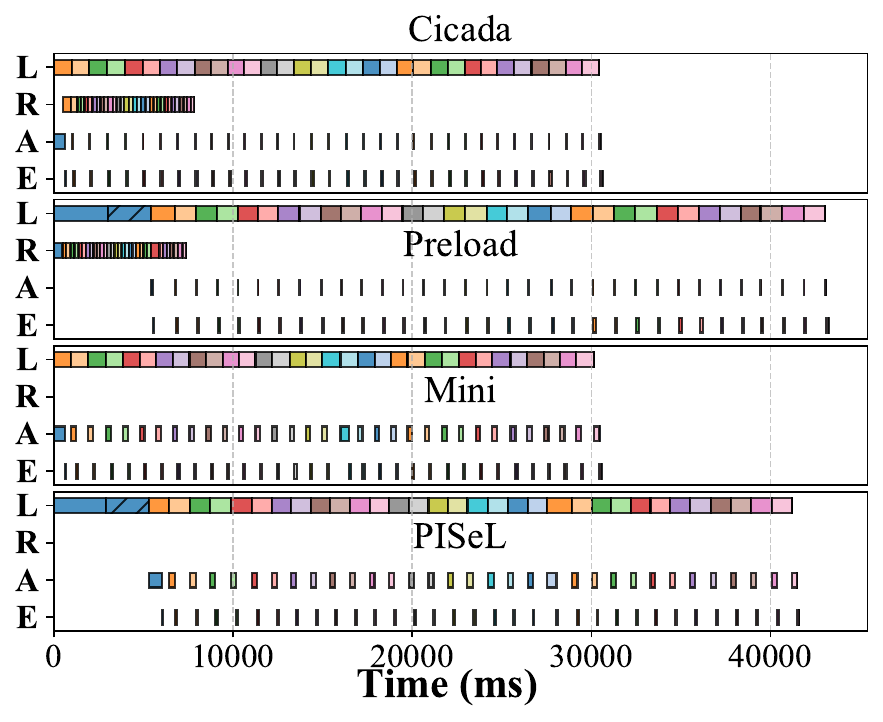}
        \label{fig:pipelineTimelineLLaMA38B}
    }
    \subfigure[OPT-6.7B]{
        \includegraphics[width=0.3\textwidth]{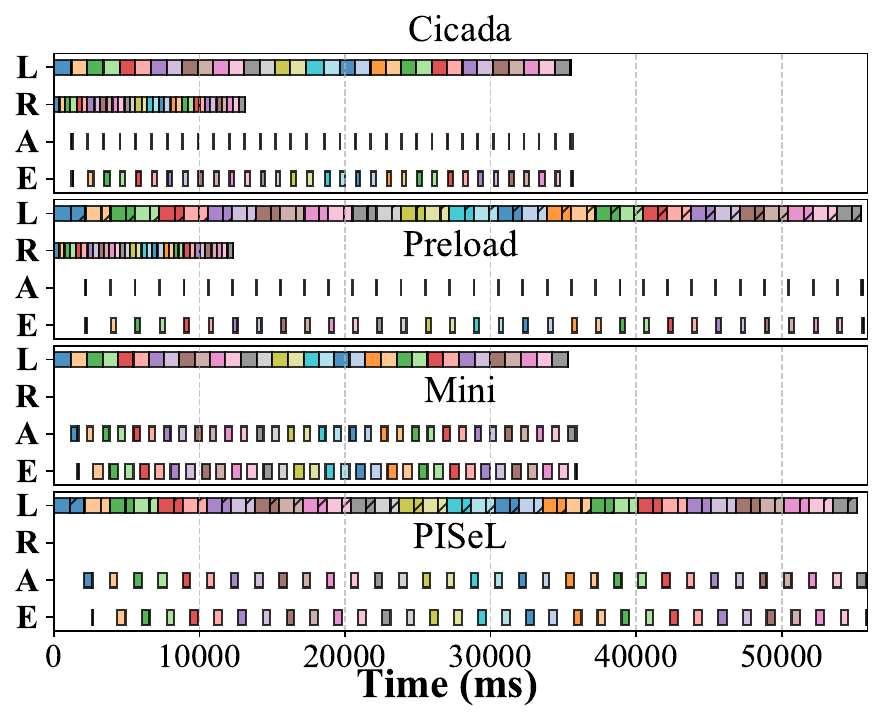}
        \label{fig:pipelineTimelineOPT67B}
    }
    \caption{Pipeline timeline for different models across strategies}
    \label{fig:pipelineOverhead}
\end{figure*}

\subsection{Evaluating Pipeline's Efficiency}
\label{sec:sub:evaluationPipelineEfficiency}
This section examines the overall efficiency of the inference pipeline by evaluating pipeline utilization and visualizing the pipeline timeline.

\subsubsection{\textbf{Pipeline Utilization}}

Pipeline utilization represents the ratio between the union of the active times of the pipeline units (after merging overlapping intervals, see Sec.~\ref{sec:evaluationMetrics} Q4) and the total pipeline execution time.
This metric provides an effective evaluation of the efficiency of resource scheduling and the degree of parallelism within the inference pipeline.

Fig.~\ref{fig:pipelineUtilization} presents the pipeline utilization for different models across PISeL, Mini, Preload and Cicada.
From the results analysis, the pipeline utilizations of PISeL and Preload are similar, while those of the Mini and Cicada strategies are also comparable. 
The primary distinction between these two groups is that the latter incorporates the MiniLoader component.
Across all evaluated models, the pipeline utilization under Mini and Cicada ranges from 99.41\% to 100\%, demonstrating consistently high efficiency.
In contrast, PISeL and Preload yield significantly lower utilization, ranging from 59.39\% to 96.89\%.
This corresponds to improvements of 31.24\%, 39.83\%, 32.66\%, and 8.17\% for ResNet, VGG, OPT, and LLaMA, respectively, yielding an average improvement of 27.98\% across the model families.
This clear gap highlights the effectiveness and robustness of the MiniLoader design in maximizing pipeline efficiency across both large language models and vision architectures.

In the following analysis, we dissect the pipeline's active and total pipeline time by categorizing the strategies into two groups: strategies with MiniLoader (Cicada) and strategies without MiniLoader (PISeL). 
This grouping is motivated by the observation that within each group, the pipeline utilization values are quite similar, allowing us to more effectively investigate the decisive contributors to pipeline utilization. 
Fig.~\ref{fig:pipelineTimesComparison} illustrates the \textbf{total active time} and \textbf{total pipeline time} for the Cicada and PISeL strategies. 
The results indicate that the disparity in pipeline utilization primarily stems from PISeL incurring a significantly longer total pipeline time compared to Cicada. 
Notably, the total pipeline time for Cicada nearly overlaps with its active time, as their values are almost identical.
In PISeL, the total pipeline time substantially exceeds the active time, ranging from 1.06x to 1.66x, suggesting that a considerable portion of the pipeline duration is idle.

It is important to note that the idle time of the pipeline does not directly equate to the sum of the waiting times for each phase.
In general, the idle time is less than the total waiting time,
Because the pipeline may still be executing other tasks while one unit is waiting.
For instance, during a weight loading task waiting for aviable resource, the layer loading or inference execution units remain active.
Moreover, as discussed in Sec.~\ref{sec:breakdownOfTimeOverheadOfPipelineUnits}, Cicada and its various implementations exhibit prolonged waiting phases relative to PISeL, longer waiting times do not necessarily imply inefficiency. 
Instead, Cicada's active time is nearly equivalent to its total pipeline time, which is the key factor enabling this strategy to achieve a pipeline utilization close to 100\%.

\subsubsection{\textbf{Pipeline Timeline}}

Fig.~\ref{fig:pipelineOverhead} illustrates the pipeline timeline (Gantt Chart) for three representative models, ResNet152, LLaMA-3-8B, and OPT-6.7B, across different strategies.
In each subfigure, the x-axis represents the processing time of the inference pipeline, and the y-axis, arranged from top to bottom, represents the pipeline stages: Layer Construction (L), Weight File Retrieval (R), Weight Application (A), and Inference Execution (E). 
Note that the Retrieve stage is only available in the Preload and Cicada strategies, which benefit from the WeightDecoupler component. 
Additionally, in the Preload and PISeL strategies, hatched bars represent the processing time of parameter registration and initialization (see Sec.~\ref{sec:constructionOverhead}). 
Moreover, different colors are used to distinguish between layers.

From Fig.~\ref{fig:pipelineOverhead}, it is evident that Cicada delivers a significant enhancement in the inference pipeline compared to other strategies.
Comparing the Preload and Cicada strategies, where the MiniLoader component is the only difference, we can see that the Layer stage is significantly reduced, therefore advancing the Retrieve, Weight and Compute start time.
Similarly, Mini reduces the Layer procedure compared to PISeL, showing the feasibility of the MiniLoader.
This advancement indicates that the MiniLoader component effectively eliminates the parameter registration and initialization phases, which are the most time-consuming operations.

To evaluate the impact of the WeightDecoupler component, we compare Mini versus Cicada and PISeL versus Preload, where the only architectural difference is the presence of WeightDecoupler.
In ResNet152 model, WeightDecoupler brings clear performance benefits by enabling the overlap between weight retrieval and layer construction, which in turn allows the Weight Application stage to be triggered immediately after the Layer stage. This effectively advances the Execution stage and improves pipeline efficiency.
For large language models such as LLaMA-3-8B and OPT-6.7B, the Layer stage dominates the overall latency, reducing the relative impact of the retrieval and application stages.
Please note that the limited performance gain is not due to inefficiency in the component design, but rather attributable to the significant structural initialization overhead inherent in LLMs.
These observations underscore the component's strength in scenarios where I/O latency constitutes a meaningful portion of the critical path, while also affirming the adaptability of Cicada's modular design across diverse model architectures.

In short, Cicada accelerates inference latency by eliminating the time-consuming parameter registration and initialization, and further overlaps the Retrieve and Layer stages, enabling immediate execution of the Weight stage and further expediting the Compute stage.

\section{Related Work}

Extensive research has focused on optimizing cold start latency in serverless computing, mainly through runtime-level optimizations~\cite{Du2020Catalyzer, Zijun2022Pagurus,Yu2024RainbowCake, Lan2024Snapipeline,Shin2022Fireworks} and application-level strategies~\cite{Stojkovic2023MXFaaS, Wu2023FaaSBatch, Wu2023HashCache, Bhasi2021Kraken, fuerst2021faascache}.

In serverless DNN inference, the cold start problem is further exacerbated by the overhead of loading large machine learning models into memory~\cite{Pei2023AsyFunc, Sui2024InstaInfer, Golec2024ColdStart, Wang2024Advancing}. 
To accelerate model inference, prior works have investigated resource reuse strategies. 
Tetris~\cite{Li2022Tetris} and Optimus~\cite{Hong2024Optimus} are two notable examples: Tetris enhances memory efficiency by sharing tensors across model-serving instances, while Optimus reduces model loading overhead by restructuring layers across inter-function invocations. 
However, both approaches introduce inherent trade-offs: Tetris necessitates a centralized tensor pool within the serverless platform, doubling memory overhead (due to tensor pool allocation alongside container memory), whereas Optimus relies on idle containers to retain model structures, increasing memory consumption. 
In contrast, Cicada optimizes layer construction by addressing inefficiencies in parameter registration and initialization. 
Through precision adjustment and selective initialization bypassing, it reduces both memory footprint and construction time, offering a more lightweight and efficient alternative.

Beyond traditional model loading optimizations, researchers have explored pipeline-based designs to enhance efficiency in Large Language Model (LLM) inference~\cite{Ma2023PipeLLM, Butler2024PipeInfer, Yu2024TwinPilots}. For instance, SplitWise~\cite{Patel2024Splitwise} partitions compute-intensive prompt processing and memory-intensive token generation across different devices, employing pipelining techniques for KV-cache transmission. ServerlessLLM~\cite{Fu2024ServerlessLLM} introduces a serverless LLM inference framework that leverages checkpointing to expedite model loading and live migration strategies to dynamically allocate checkpoints across devices, maximizing serverless scalability. While these approaches primarily focus on optimizing resource allocation and scheduling during inference, Cicada complements them by specifically improving model loading efficiency, contributing to a more comprehensive optimization strategy.

Several studies have also investigated pipelined model loading for improved efficiency. DistMind~\cite{Jin2024DistMind} and Demand Layering~\cite{Ji2022DemandLayering} partition the traditionally sequential model loading process into three stages, leveraging SSD hardware and zero-copy techniques for weight file retrieval. Similarly, PISeL~\cite{Rahimi2024PISeL} introduces a three-stage pipeline tailored for serverless environments. However, existing pipeline designs overlook two critical aspects: (1) the unnecessary overhead incurred during layer construction in inference scenarios and (2) the feasibility of decoupling I/O operations from computational tasks in the weight loading phase. Cicada addresses these limitations through MiniLoader and WeightDecoupler components, respectively, pushing the frontier of model loading optimization.

\section{Conclusions}
This paper presents Cicada, a pipeline optimization framework for serverless DNN inference. 
By systematically analyzing the bottlenecks in existing pipeline-based model loading, Cicada introduces innovative optimizations across computational, storage, and scheduling dimensions. 
First, the MiniLoader component significantly reduces layer construction overhead through a lightweight parameter initialization strategy. 
Second, the WeightDecoupler decouples weight file retrieval from layer construction, enabling asynchronous weight prefetching and out-of-order weight application. 
Finally, the priority-aware scheduler dynamically allocates resources to ensure timely execution of high-priority inference tasks. 
Experimental results demonstrate that Cicada achieves substantial improvements in inference latency and pipeline utilization, effectively addressing the performance bottlenecks in traditional pipeline-based model loading. 


\bibliographystyle{IEEEtran}
\bibliography{Cicada.bib}


\begin{IEEEbiography}[{\includegraphics[width=1in,height=1.25in,clip,keepaspectratio]{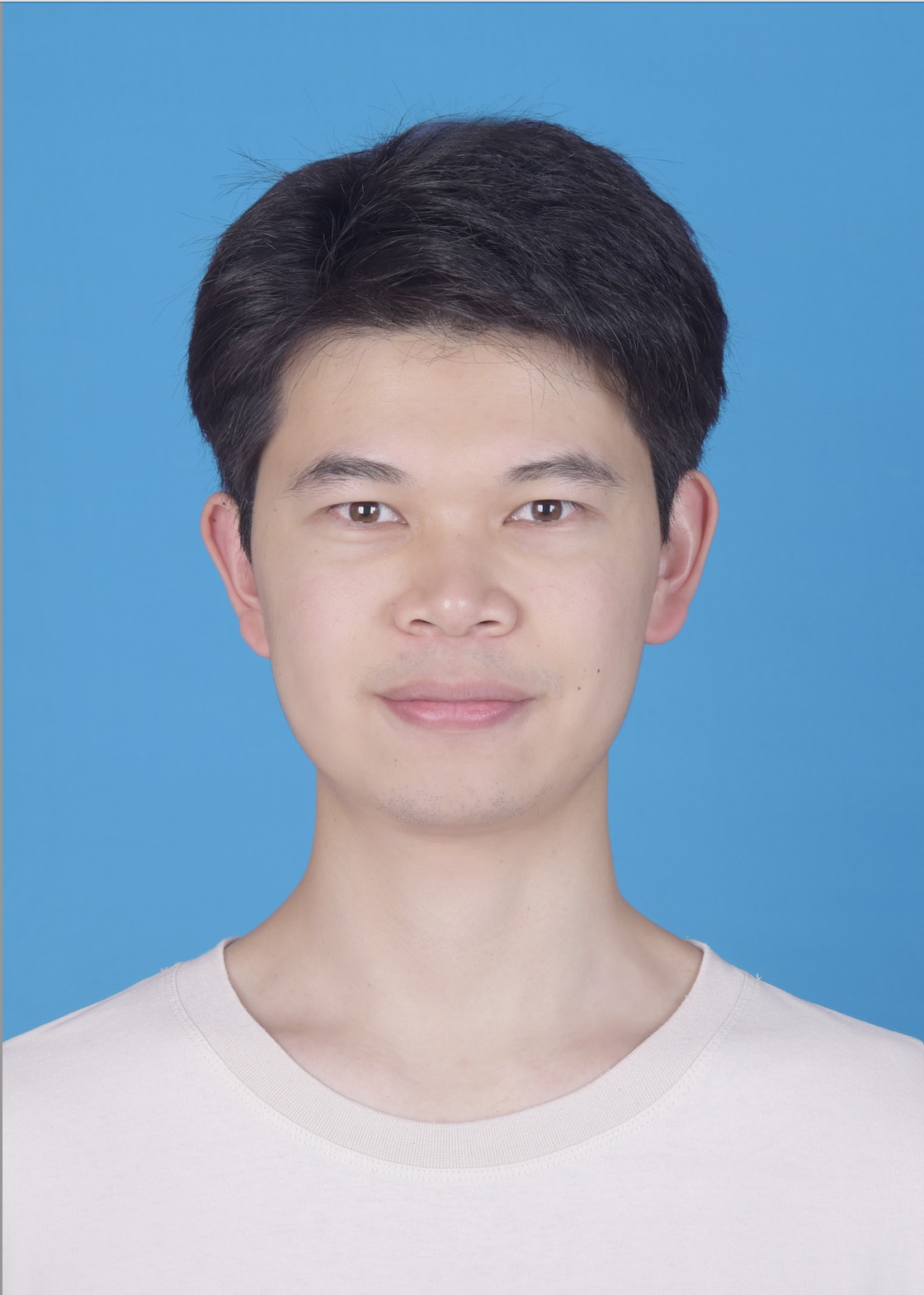}}]{Zhaorui Wu}
	is a PhD candidate of Computer Science Department, Jinan University, China and a visiting student at the University of Exeter, United Kingdom. He received the MS degree in computer architecture from the Computer Science Department of Jinan University. He has published several papers in IEEE TC, IEEE TPDS, IEEE TNSM, DATE and ICDCS. His current research interests include serverless computing, computer architecture, performance evaluation, etc.
  \end{IEEEbiography}
\begin{IEEEbiography}[{\includegraphics[width=1in,height=1.25in,clip,keepaspectratio]{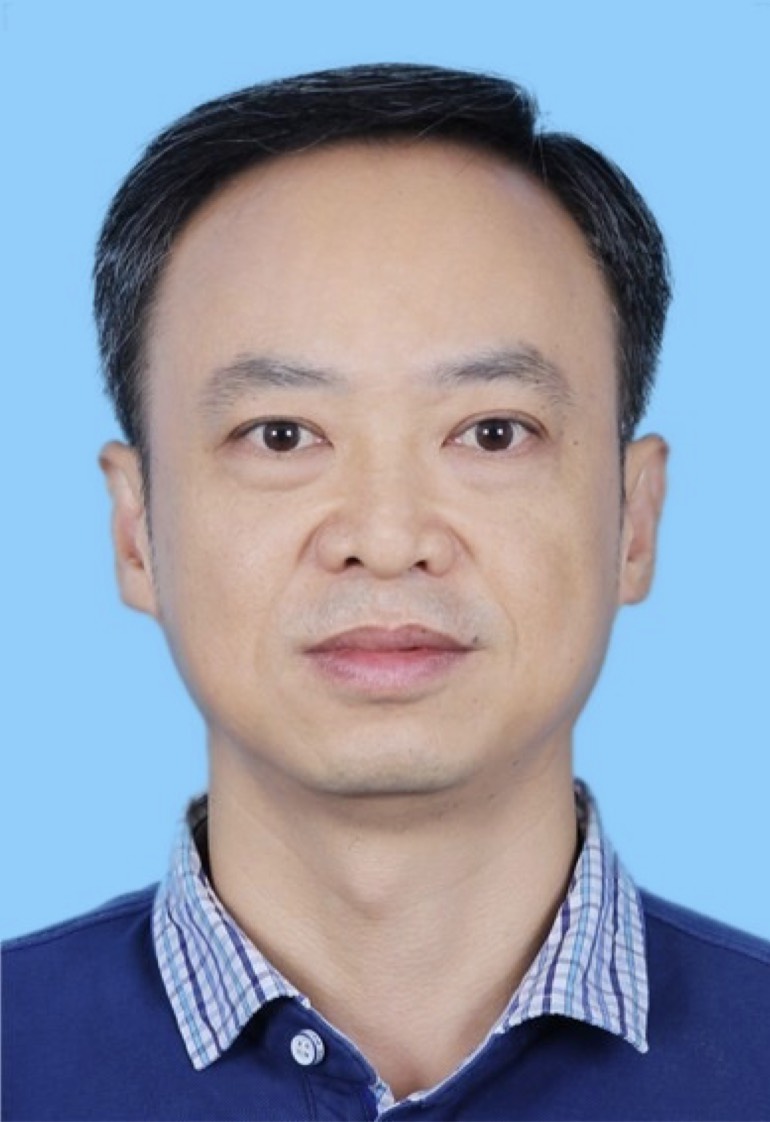}}]{Yuhui Deng}
	received the PhD degree in computer science from the Huazhong University of Science and Technology, in 2004. He is a professor with the Computer Science Department, Jinan University. Before joining Jinan University, he worked at EMC Corporation as a senior research scientist from 2008 to 2009. He worked as a research officer at Cranfield University in the United Kingdom from 2005 to 2008. He in on the finallist of storage challenge in ACM/IEEE SC2007, and on the finallist of EMC global innovation showcase in 2008. He was awarded the Annual Best Paper Award from JISE in 2017, and Best Journal Paper Award from the Big Data Technical Committee of IEEE Communications Society in 2019. His research interests cover computer architecture, cloud computing, information storage, etc.
\end{IEEEbiography}
\begin{IEEEbiography}[{\includegraphics[width=1in,height=1.25in,clip,keepaspectratio]{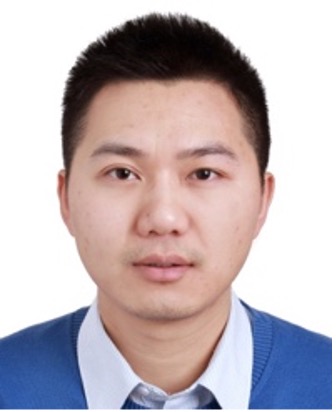}}]{Jia Hu}
	is an Associate Professor in Computer Science at the University of Exeter. He received his Ph.D. degree in Computer Science from the University of Bradford, UK, in 2010, and M.Eng. and B.Eng. degrees in Electronic Engineering from Huazhong University of Science and Technology, China, in 2006 and 2004, respectively. His research interests include edge-cloud computing, resource optimization, applied machine learning, and network security.
\end{IEEEbiography}
\begin{IEEEbiography}[{\includegraphics[width=1in,height=1.25in,clip,keepaspectratio]{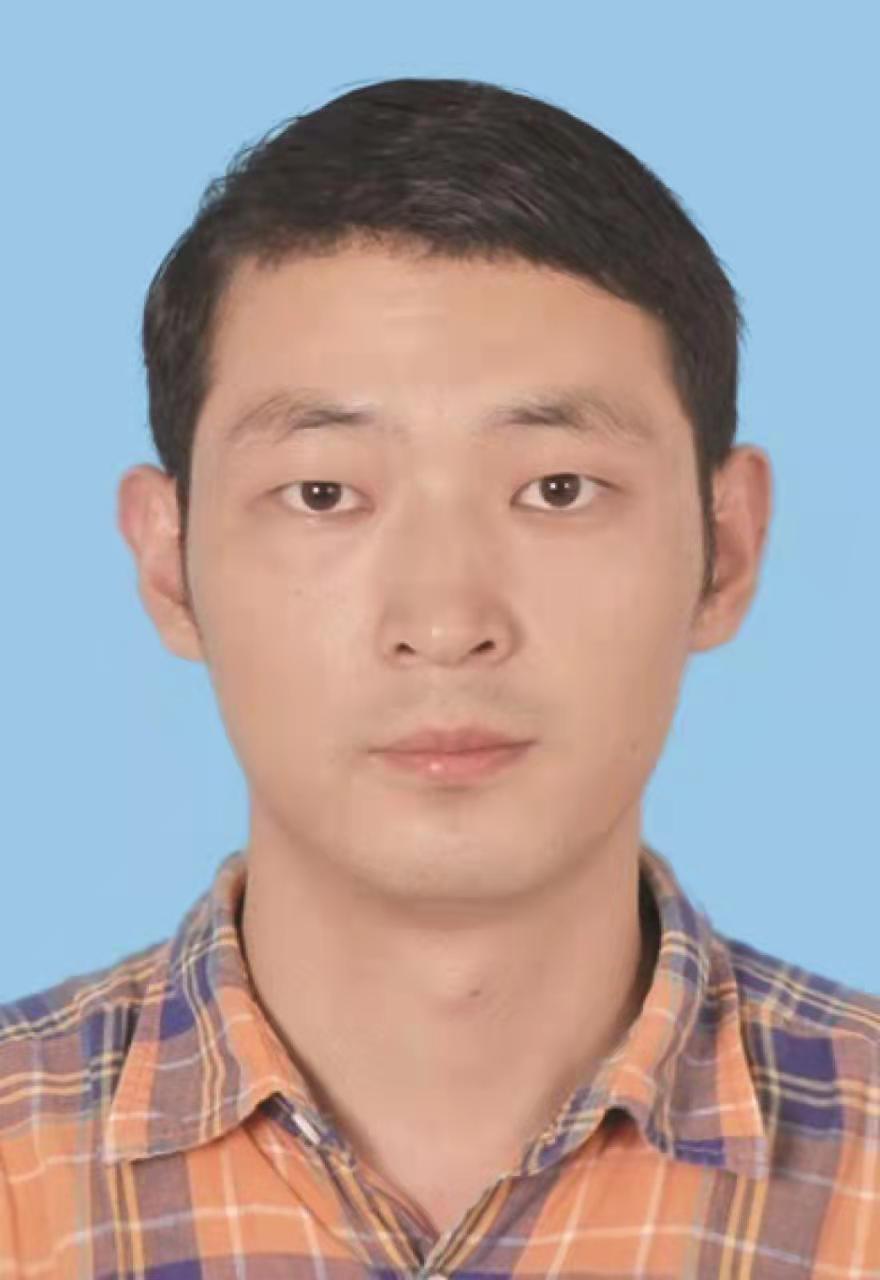}}]{Lin Cui} 
	received the Ph.D. degree from City University of Hong Kong in 2013. He is currently a professor in the Department of Computer Science at Jinan University, Guangzhou, China. He has broad interests in networking and distributed systems, with focuses on software defined networking (SDN), programmable data plane, network function virtualization (NFV), congestion control and so on.

\end{IEEEbiography}
\begin{IEEEbiography}[{\includegraphics[width=1in,height=1.25in,clip,keepaspectratio]{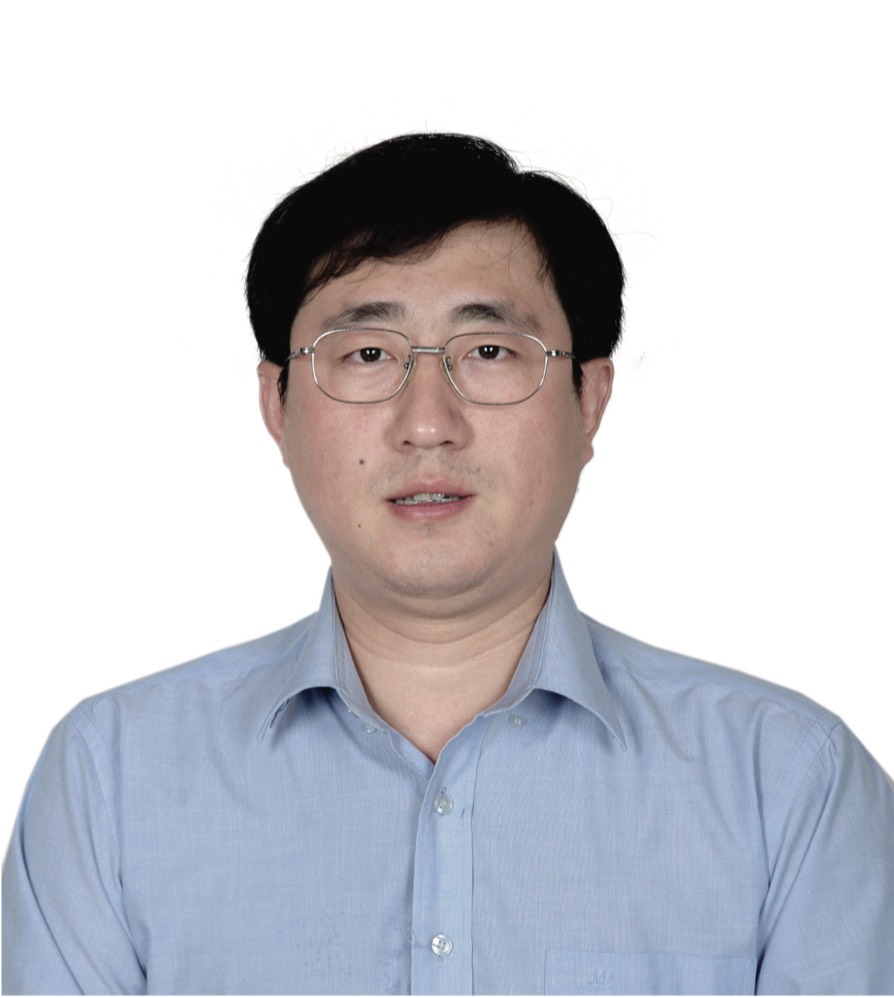}}]{Zhen Zhang}
	received the BS and MS degrees in computer science from Jilin University, China, in 1999 and 2003, respectively, and the PhD degree from the College of Computer Science and Engineering, South China University of Technology, China, in 2011. He is a professor with the Department of Computer Science, Jinan University. His research interests include graph theory, parallel and distributed processing, cloud computing and complex networks.
	\end{IEEEbiography}
\begin{IEEEbiography}[{\includegraphics[width=1in,height=1.25in,clip,keepaspectratio]{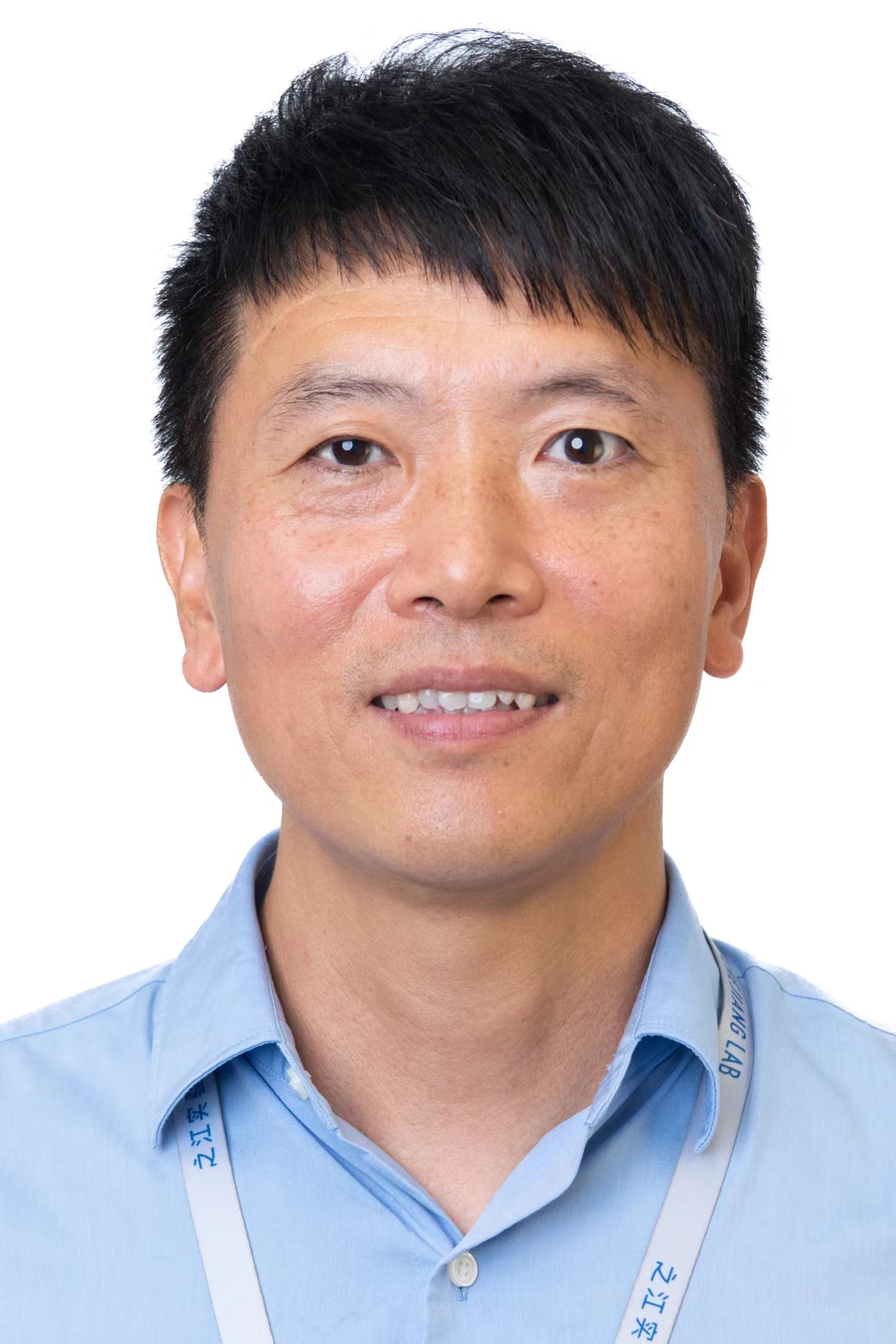}}]{Lingfang Zeng} is a Full Professor at Zhejiang Lab and an Adjunct Professor at Zhejiang University. He received the PhD degree in computer architecture from Huazhong University of Science and Technology (HUST) in 2006. He has authored over 100 papers in leading conferences and journals, including SIGMOD, FAST, SC, ACL, TPDS, TC, TKDE, TDSC, TIFS, ToS, and TACO. His contributions also include more than 100 granted patents, one book, and two book chapters, as well as participation in the development of seven industry standards. He has received several research awards, including the ACM/IEEE SC06 Storage Challenge Finalist and IEEE Best Paper Awards. He is a Distinguished Speaker and Board Member of the China Computer Federation (CCF), and a member of IEEE and ACM. He currently serves on the editorial board of the Journal of High-Performance Storage (JHPS).
\end{IEEEbiography}
	
\begin{IEEEbiography}[{\includegraphics[width=1in,height=1.25in,clip,keepaspectratio]{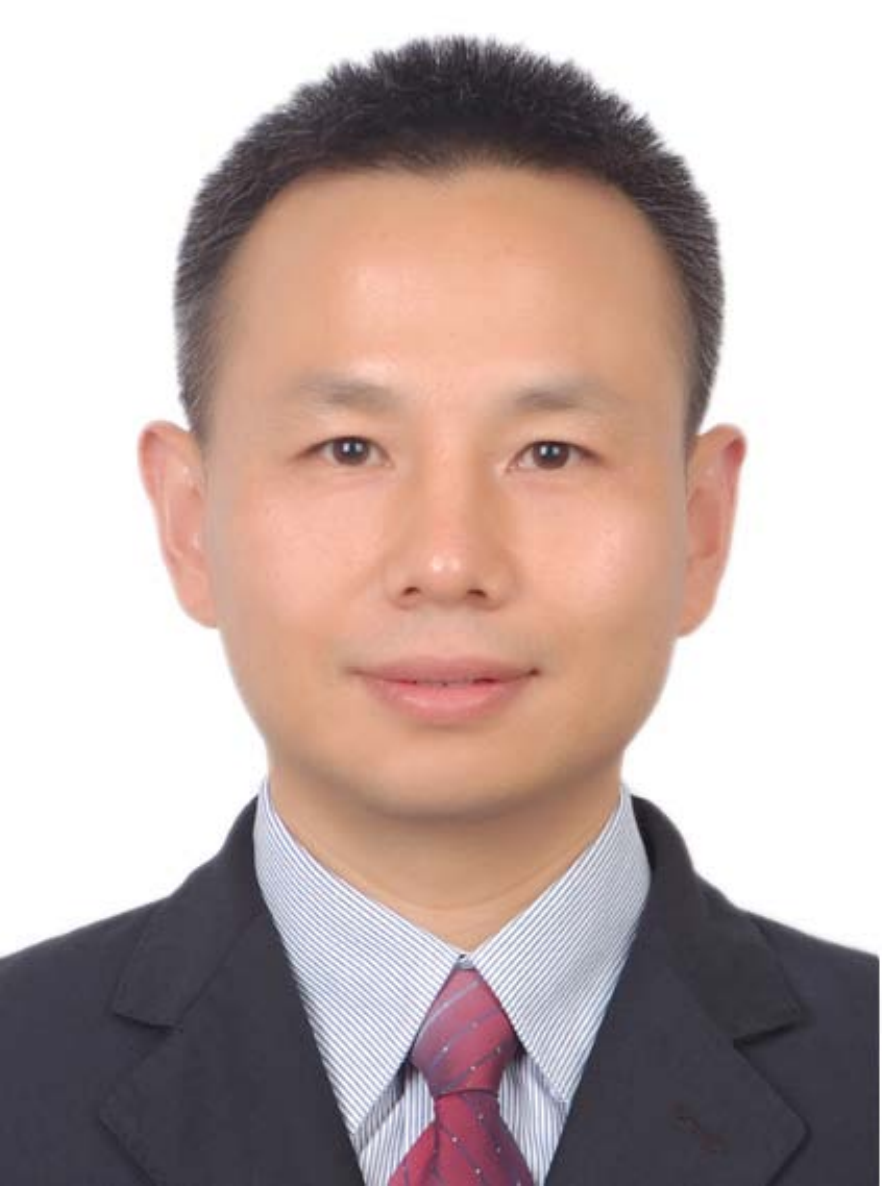}}]{Geyong Min} is a Professor of High Performance Computing and Networking in the Department of Mathematics and Computer Science within the College of Engineering, Mathematics and Physical Sciences at the University of Exeter, United Kingdom. He received the PhD degree in Computing Science from the University of Glasgow, United Kingdom, in 2003, and the B.Sc. degree in Computer Science from Huazhong University of Science and Technology, China, in 1995. His research interests include Future Internet, Computer Networks, Wireless Communications, Multimedia Systems, Information Security, High Performance Computing, Ubiquitous Computing, Modelling and Performance Engineering. He served as an Associate Editor of IEEE Transactions on Cloud Computing, IEEE Transactions on Sustainable Computing, IEEE Transactions on Computers.
\end{IEEEbiography}
\end{document}

%% file: abstract.tex
\begin{abstract}
Serverless computing has emerged as a pivotal paradigm for deploying Deep Neural Network (DNN) models, including both large language models (LLMs) and foundation models designed for a wide range of tasks.
However, addressing the overlooked gap of time-intensive model loading in serverless DNN inference systems remains challenging, since the time-intensive construction of layers and the monolithic loading of weights contribute to suboptimal resource utilization. 
In this paper, we propose \textit{Cicada}, a novel pipeline optimization framework for DNN inference that leverages a resource-centric architecture to mitigate structural overhead and pipeline stalls. 
Cicada introduces two key innovations: MiniLoader and WeightDecoupler. MiniLoader eliminates redundant parameter initialization by exploiting low-precision representations during layer construction, reducing structural overhead. In parallel, WeightDecoupler decouples weight retrieval from its application, enabling asynchronous weight fetching and out-of-order weight application to remove pipeline stalls.
Our experimental results demonstrate that Cicada outperforms the state-of-the-art frameworks.
Specifically, Cicada reduces inference latency by an average of 34.79\%, with the MiniLoader component contributing the majority of this optimization (up to 47.34\%), and the WeightDecoupler achieves up to 23.87\% improvement. 
Additionally, Cicada achieves up to 68.49\% speedup in the inference pipeline utilization, reaching nearly 100\% overall occupancy.
Moreover, Cicada reduces memory usage by 45.77\% to 81.82\% compared to state-of-the-art strategies.
\end{abstract}